\documentclass[sn-basic,pdflatex]{sn-jnl}

\usepackage{graphicx}%
\usepackage{multirow}%
\usepackage{amsmath,amssymb,amsfonts}%
\usepackage{amsthm}%
\usepackage{mathrsfs}%
\usepackage[title]{appendix}%
\usepackage{textcomp}%
\usepackage{manyfoot}%
\usepackage{booktabs}%
\usepackage{algorithm}%
\usepackage{algorithmicx}%
\usepackage{algpseudocode}%
\usepackage{listings}%
\usepackage{xcolor}
\definecolor{mylightgreen}{HTML}{4FADA5}

\usepackage{times}
\usepackage{soul}
\usepackage{url}
\usepackage[subrefformat=parens]{subcaption}
\usepackage{float}
\usepackage{tikz}

\usepackage{longtable}
\usepackage{multirow}
\usepackage{makecell}

\definecolor{myorange}{RGB}{255, 102, 0}
\definecolor{mygreen}{RGB}{0, 176, 165}

\usepackage{bbold}
\usepackage[export]{adjustbox}
\usepackage{lscape}

\usepackage{comment}

\newcommand{\argmin}[2]{\mathrm{arg}\;\underset{#1}{\min}\;#2}

\newcommand{\Mycomment}[1]{}

\theoremstyle{thmstyleone}%
\theoremstyle{thmstyletwo}%

\theoremstyle{thmstylethree}%

\begin{document}

\title[Article Title]{An Inference-Based Architecture for Intent and Affordance Saturation in Decision-Making}


\author*[1]{\fnm{Wendyam Eric Lionel} \sur{Ilboudo}}\email{ilboudo.wendyam\_eric.in1@is.naist.jp}

\author*[1]{\fnm{Saori C} \sur{Tanaka}}\email{x.saori@is.naist.jp}

\affil*[1]{\orgdiv{Division of Information Science}, \orgname{Nara Institute of Science and Technology}, \orgaddress{\street{8916-5 Takayama-cho}, \city{Ikoma}, \postcode{630-0192}, \state{Nara}, \country{Japan}}}


\abstract{
Decision paralysis, i.e. hesitation, freezing, or failure to act despite full knowledge and motivation, poses a challenge for choice models that assume options are already specified and readily comparable. Drawing on qualitative reports in autism research that are especially salient, we propose a computational account in which paralysis arises from convergence failure in a hierarchical decision process. We separate intent selection (what to pursue) from affordance selection (how to pursue the goal) and formalize commitment as inference under a mixture of reverse- and forward-Kullback--Leibler (KL) objectives. Reverse KL is mode-seeking and promotes rapid commitment, whereas forward KL is mode-covering and preserves multiple plausible goals or actions. In static and dynamic (drift–diffusion) models, forward-KL-biased inference yields slow, heavy-tailed response times and two distinct failure modes, intent saturation and affordance saturation, when values are similar. Simulations in multi-option tasks reproduce key features of decision inertia and shutdown, treating autism as an extreme regime of a general, inference-based, decision-making continuum.
}





\maketitle


\section{Introduction}

Human decision-making depends on continuous transformation of perceptual input into stable intentions and actionable affordances. Although this process typically operates smoothly, even neurotypical individuals experience moments of decision paralysis, i.e., hesitation in the face of ambiguous tasks, transient freezing under multitasking pressure, or difficulty initiating action despite clear goals. Autistic reports provide an amplified, theoretically informative expression of these disruptions. Phenomena such as autistic inertia (difficulty initiating, stopping, or shifting action and autistic shutdown (overload-induced collapse of responsiveness) illustrate how action formation can fail when competing internal demands cannot be resolved. First-person descriptions further refine this picture through two experiential categories \citep{ayaya2022asd}: \emph{intent saturation}, in which perceptual and semantic information accumulate without yielding a clear intention about \emph{what} to do, and \emph{affordance saturation}, in which numerous potential actions become simultaneously salient, overwhelming the system's ability to determine \emph{how} to act. These states can be understood as phenomenological subdivisions of inertia and shutdown, respectively, revealing distinct stages at which perception-action coupling may falter. Importantly, these mechanisms are not unique to autism. Instead, they represent intensified expressions of vulnerabilities present in all human decision systems. Autistic experiences therefore offer a magnified view of the computational principles, such as filtering, prioritization, and competition among action representations, which support emergence of intentions and actions, but which can break down in both typical and atypical cognition.

Despite the ubiquity of such phenomena, current theories do not provide a unified account of how intention formation and affordance selection become unstable under conditions of overload. In behavioral neuroscience, decision-making is typically framed as comprising distinct valuation and comparison stages \citep{krajbich2010visual, wallis2007orbitofrontal}. Consequently, classical models focus on value comparison or evidence accumulation among options   that are already specified, leaving unexamined the upstream processes by which intentions and candidate actions are generated and prioritized. Conversely, accounts of executive dysfunction or cognitive control emphasize downstream regulatory failures, but shed little light on how perceptual and other relevant information can accumulate in ways that destabilize formation of a coherent action plan. Autistic narratives, precisely because they reveal early, fragile points of this process, indicate that paralysis can arise not only from conflict between evaluated alternative actions, but also from breakdowns in generation and filtering of possible goals and action paths. Yet existing computational frameworks do not formalize these distinct points of failure or explain how they interact to produce hesitation, freezing, or shutdown across different contexts. A model that explicitly distinguishes intention formation, affordance extraction, and action commitment, and that specifies mechanisms by which these stages can saturate or fail to converge, would therefore extend our understanding of decision formation in both typical and atypical minds.

In that spirit, the present paper introduces a computationally explicit framework that explains decision paralysis in terms of failures in intent and affordance selection. Our first hypothesis is that \emph{intent saturation} (corresponding to inertia) and \emph{affordance saturation} (corresponding to shutdown-like overload) arise when multiple options are represented as comparably valuable at two distinct levels of a hierarchical decision process. At the \emph{intent level}, the agent must select what outcome or goal to pursue. At the \emph{affordance level}, the agent must determine how to realize that intent through concrete actions offered by the environment. To capture both points of saturation, we model decision-making as involving an explicit stage of intent selection preceding action selection, rather than collapsing both into a single choice over actions, as in standard reinforcement-learning formulations. Our second hypothesis concerns the divergence measure that governs how internal policies approximate a target distribution over options. Much of computational neuroscience implicitly relies on minimizing the reverse Kullback-Leibler (RKL) divergence, which is mode seeking, whereas the forward KL (FKL) divergence is mean seeking and tends to allocate probability mass across multiple modes.
 We propose that decision-making typically dominated by RKL-like objectives, encourages sharp commitment to a single option, whereas decision-making more strongly influenced by FKL-like objectives, encourages broader coverage of multiple possibilities.
  With this view, decision paralysis emerges when an FKL-biased system encounters many similarly plausible goals or actions, producing an overinclusive, multimodal representation that overwhelms the mechanisms required for commitment.

\section{Results}

In this study, we test two hypotheses derived from the two-stage structure of decision-making outlined above. The first hypothesis is that decision paralysis arises when comparison fails to converge at either the intent level (what outcome to pursue) or the affordance level (how to achieve it), particularly when several options have similar value. Our model therefore incorporates both intention selection and action selection.
The second hypothesis is that the nature of this breakdown depends on the divergence principle guiding inference. Whereas models based on the reverse KL (RKL) tend to force commitment to a single option, forward KL (FKL)-dominated inference preserves multiple possibilities, leading to non-convergent comparison and cognitive paralysis.

To examine these hypotheses, we integrate them into a coherent decision-making model (see Figure~\ref{fig:graphical_abstract} for an overview). In this section, we present results for a static version of the model which serves as a counterpart to the conventional softmax-based choice model with no time dependence, and for a dynamic version corresponding to evidence accumulation, implemented as a drift-diffusion process with KL-based drift and a decision reached when the time-dependent variable $x_t$ hits a boundary (details in Methods and Supplementary Information).

\subsection{Simulation Scenarios}

To illustrate how decision paralysis and slow decision-making arise in the proposed model, we simulate two representative tasks. During the decision process, the agent first computes the state-action value functions using dynamic programming, e.g., value iteration~\cite{sutton2018reinforcement}. Throughout all simulations, we assume that the agent has access to the task structure and can therefore evaluate these values exactly. This assumption is not restrictive. Decision paralysis frequently occurs in everyday situations in which the environment is already familiar. The resulting value distributions determine both the optimal intent and associated actions, represented as a discrete Boltzmann (softmax) distribution, which then governs action selection.

\subsubsection{RGB Token One-Step Task}

\begin{figure*}[b]
    \centering
    \captionsetup{justification=centering}
    \begin{minipage}[c]{0.5\linewidth}
        \centering
        \includegraphics[width=\linewidth]{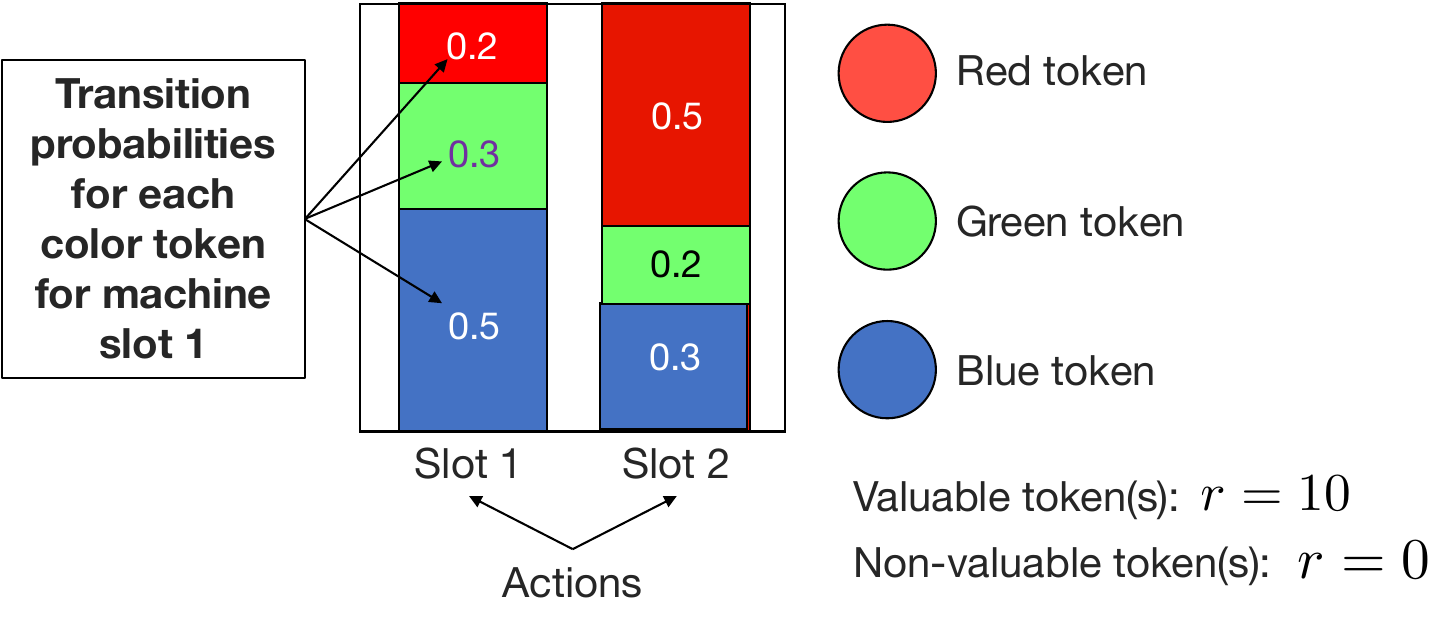}
        \subcaption{One-step decision task (RGB token world).}
        \label{fig:rgb_task}
    \end{minipage}
    \hspace{0.05\linewidth}
    \begin{minipage}[c]{0.35\linewidth}
        \centering
        \includegraphics[width=\linewidth]{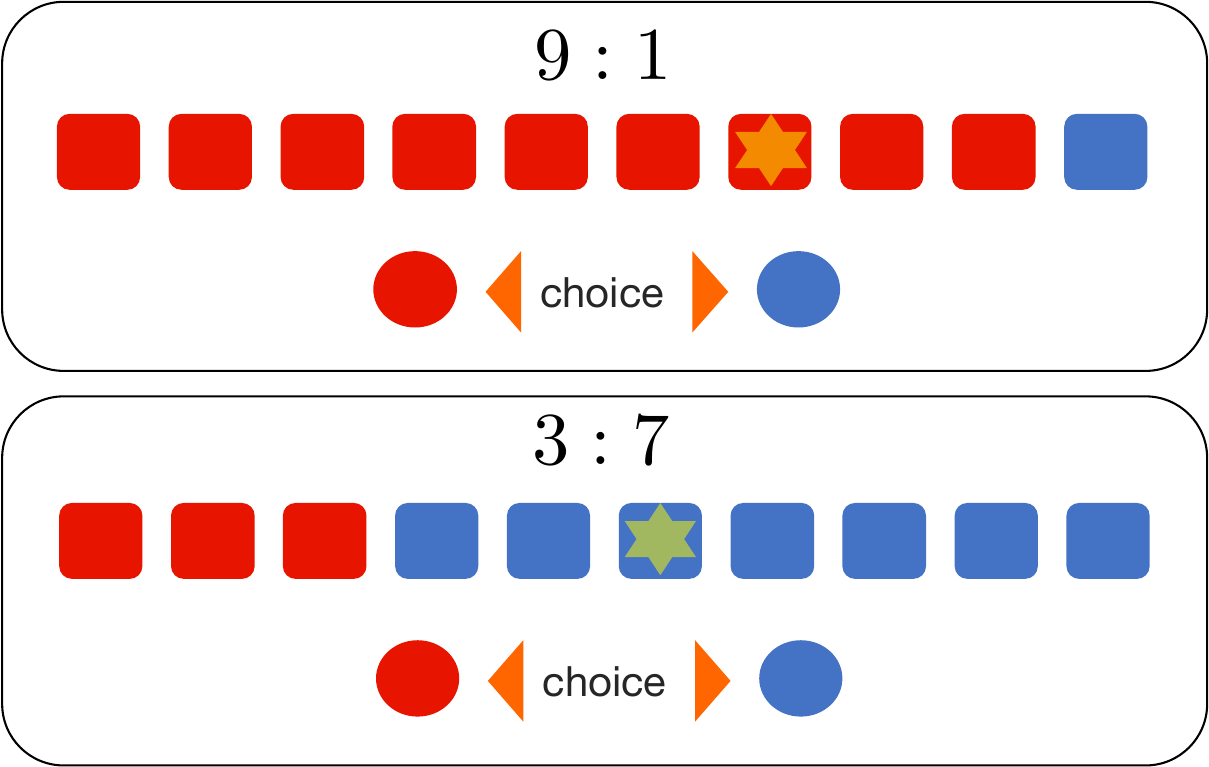}
        \subcaption{Cambridge Gamble Task (CGT).}
        \label{fig:cgt_task}
    \end{minipage}
    \caption{Tasks used to evaluate the proposed KL-based decision model.}
    \label{fig:sim_tasks}
\end{figure*}

This one-step decision task, adapted from the observational learning paradigm in~\cite{wu2024individual}, requires the agent to choose one of several slot machines (Fig.~\ref{fig:rgb_task}). Each machine yields a red, green, or blue token with fixed probabilities. Depending on the structure of the task, the agent may face ambiguity at the level of intent (which token to seek for) or affordance (which action best yields that token). We examine four cases:

\begin{enumerate}
    \item \textbf{Case 1:} One valuable token and one optimal action (fully unambiguous).
    \item \textbf{Case 2:} One valuable token, but multiple optimal actions.
    \item \textbf{Case 3/4:} Multiple valuable tokens, each with a unique optimal action. Case 3 has a smaller value differences than Case 4.
\end{enumerate}

These conditions span situations with clear or competing intents (valuable tokens) and clear or competing affordances (actions). Case 1 represents an environment with a single dominant goal and action. Case 2 introduces action-level ambiguity. Cases 3 and 4 introduce ambiguity at the level of goal selection, with different degrees of separation between the available options.

\subsubsection{Cambridge Gamble Task (First Stage)}

The first stage of the Cambridge Gamble Task (CGT~\cite{rogers1999dissociable}) provides a natural two-choice analogue to the RGB token task. Participants are shown a row of ten boxes, some red and some blue, one of which hides a token. The ratio of red to blue boxes varies across trials (from 9:1 to 5:5). The agent must select whether the token is more likely to be under a red or blue box.

Computationally, this first stage corresponds to a two-color version of the RGB task, where the box ratio directly determines the value of each action. We therefore simulate only this stage, as it directly reflects decision time and ambiguity without involving learning or reward weighting.

\subsection{Results of Computational Simulations}

\subsubsection{RGB Token One-Step Task Results}

 Results for the static model are given in Fig.~\ref{fig:rgb_static_res}. In the static model, the decision variable is not dependent on time. Instead a decision is made if the sampled variable falls within a region of decidability. In Fig.~\ref{fig:rgb_static_res}, this static decision is captured by the arrow that shows the most likely position for the decision variable. As can be seen, differences occurs depending on the task condition and on the specific KL model involved in the decision process.
 Specifically, in Case 1, when the task is unambiguous, with a single clear optimal token (intent = ``get green token") and a clear corresponding optimal action (``select slot 1 to maximize chance of getting green token"), no qualitative difference exists between the FKL and RKL models, as both correctly select the green token as their goal and slot 1 as their action. Hence, no saturation occurs.
 In contrast, in Case 2 where the single optimal token can be obtained through both slot machines, giving them the same value, the FKL model's most likely position falls within a region of the decision space corresponding to neither of the slots, leading to a saturation of affordance where although the agent knows what it wants to achieve, the tug of war between the two actions overwhelms its decision process.
 Similarly, in Case 3, where multiple optimal tokens are available, the agent must choose between two goals. However, since both goals have the same value, the agent is unable to choose between them. This renders it unable to consider what actions to take (blank). This corresponds to a saturation of intent in which the agent cannot initiate any action.
 
\begin{figure*}[b]
    \centering
    \includegraphics[width=\linewidth]{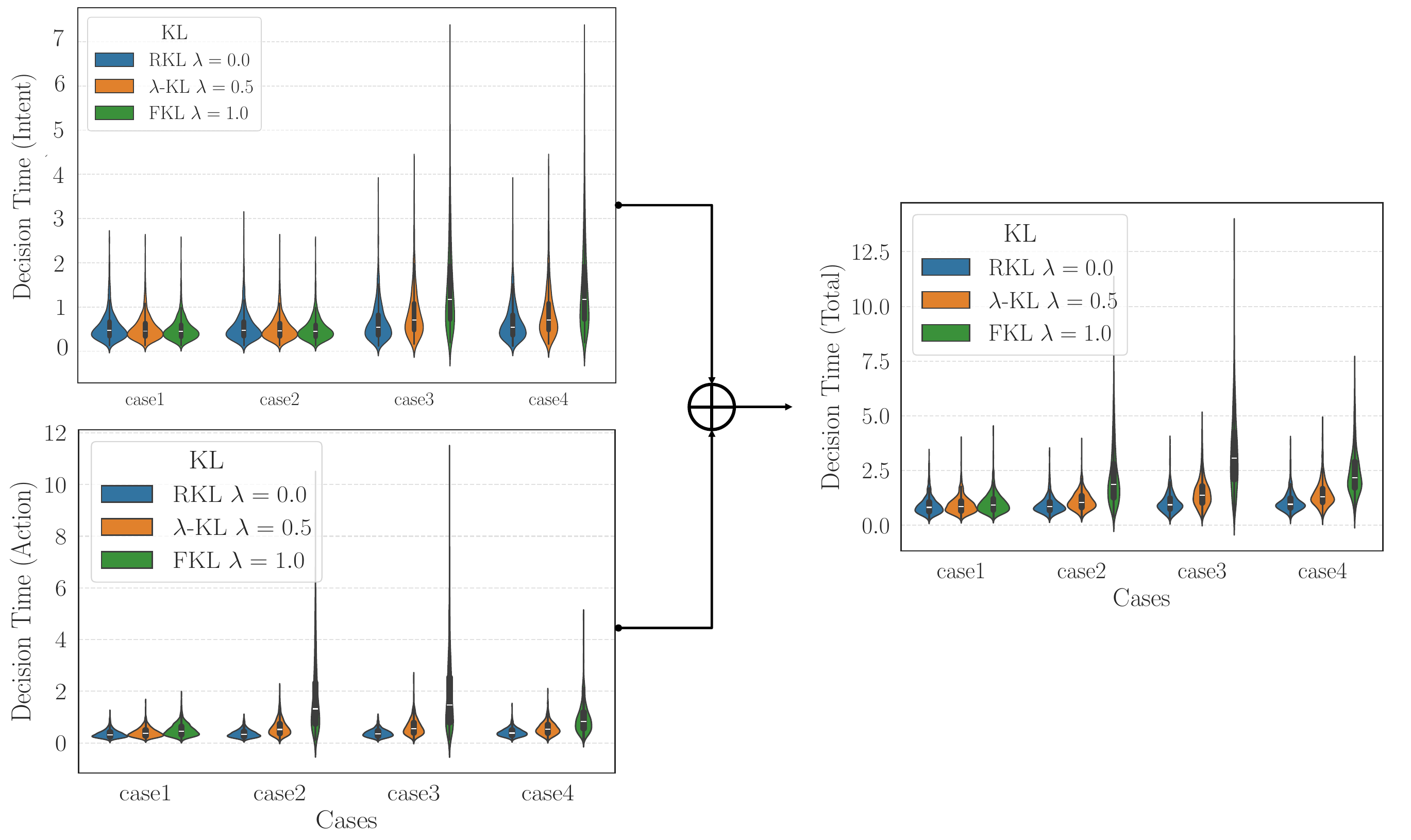}
    \caption{Violin plots of decision times for intent, action, and total decision time across all cases. The FKL regime ($\lambda=1$) exhibits consistently longer latencies, especially when multiple valuable options are available (Cases~2-4).}
    \label{fig:rgb_res_violin}
\end{figure*}

\begin{figure*}[b]
    \centering
    \includegraphics[width=\linewidth]{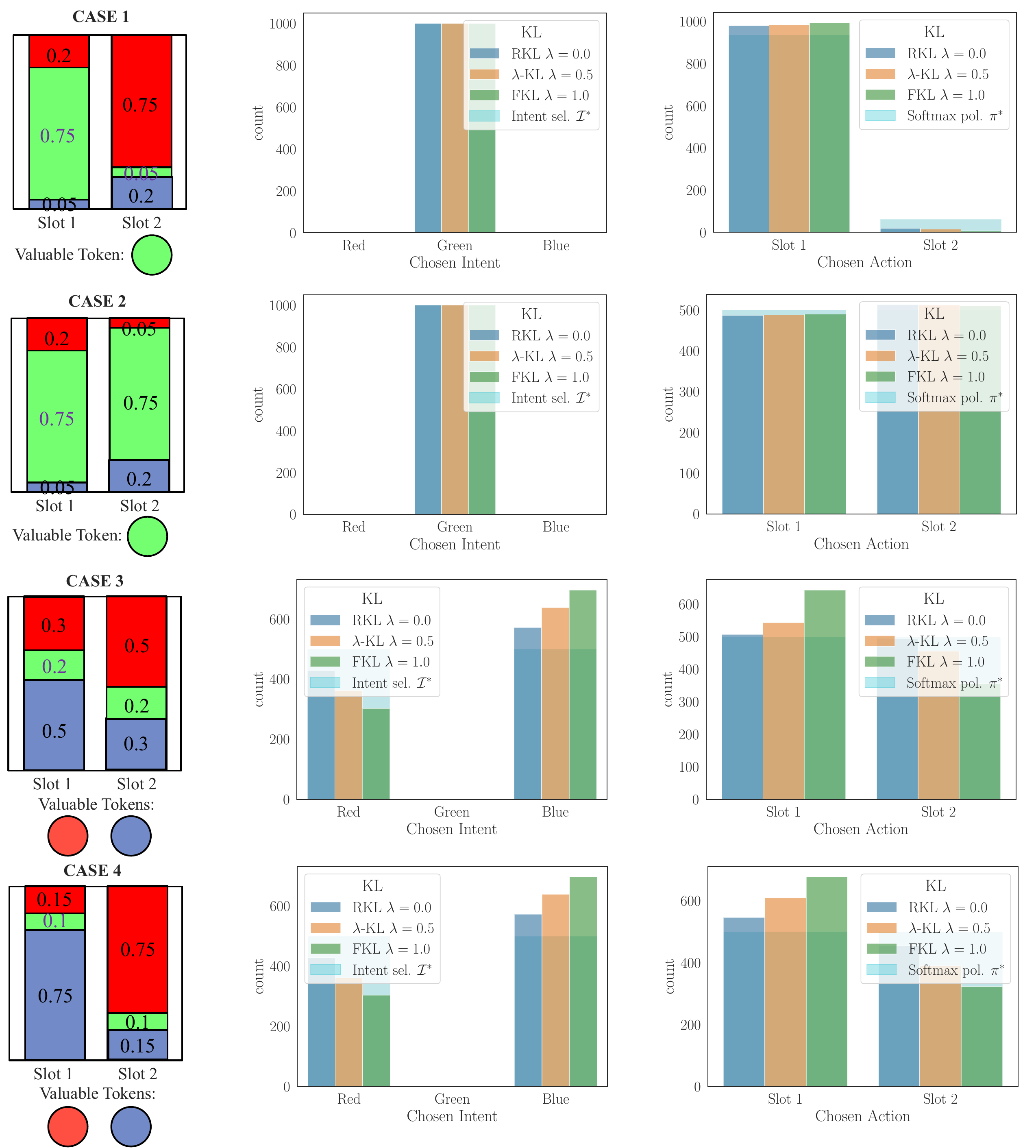}
    \caption{Selection proportions for intents and actions across 1000 drift-diffusion simulations. Intent selections match the optimal intent distribution, and action selection follows the softmax policy induced by the KL-based decision model.}
    \label{fig:rgb_selection_proba}
\end{figure*}

 In turn, figures~\ref{fig:rgb_res}, \ref{fig:rgb_res_violin}, \ref{fig:rgb_selection_proba} and \ref{fig:rgb_ddm_res} show the results of the dynamical version of the model in which the decision variable is time-dependent and evolves following a drift-diffusion equation with a drift that depends on the type of KL divergence involved. Specifically, Fig.~\ref{fig:rgb_res} shows the distribution of decision times over 1000 trials for the decision model based on the FKL, the RKL and a mixture of both (with equal weight $\lambda=0.5$). As can be seen, in cases 2 to 4 in which there are many possible valid options, the FKL and $\lambda$-KL models show distributions with heavier-tails indicating a tendency toward longer decision times. As expected, as $\lambda$ increases (tending toward an FKL-based decision), the longer it takes to make a decision. This can be viewed in the Fig.~\ref{fig:rgb_res_violin} where we also notice that the speed of the $\lambda$-KL and FKL models compared with the RKL model depends on the task structure, with no significant difference between the three models in Case 1. Fig.~\ref{fig:rgb_ddm_res} presents an example of decision variable dynamics for Case 2. As observed, all three models take approximately the same time to reach a decision on intent selection,(i.e., ``to get a green token") and differ only in the action decision, with the FKL model taking five seconds to reach a decision when the RKL model decides in less than a second. Clearly, from the perspective of the RKL-based agent, the FKL-based agent appears effectively paralyzed or indecisive. Finally, Fig.~\ref{fig:rgb_selection_proba} provides a plot of the choice distribution compared to the optimal softmax-based decision distribution, revealing that the KL-based drift-diffusion reproduces choice distributions that are consistent with a static conventional selection mechanism, despite also providing a basis for decision paralysis and slow decision through an FKL-biased decision-making processes.

\subsubsection{Cambridge Gamble Task Results}

\begin{figure*}[b]
    \centering
    \includegraphics[width=\linewidth]{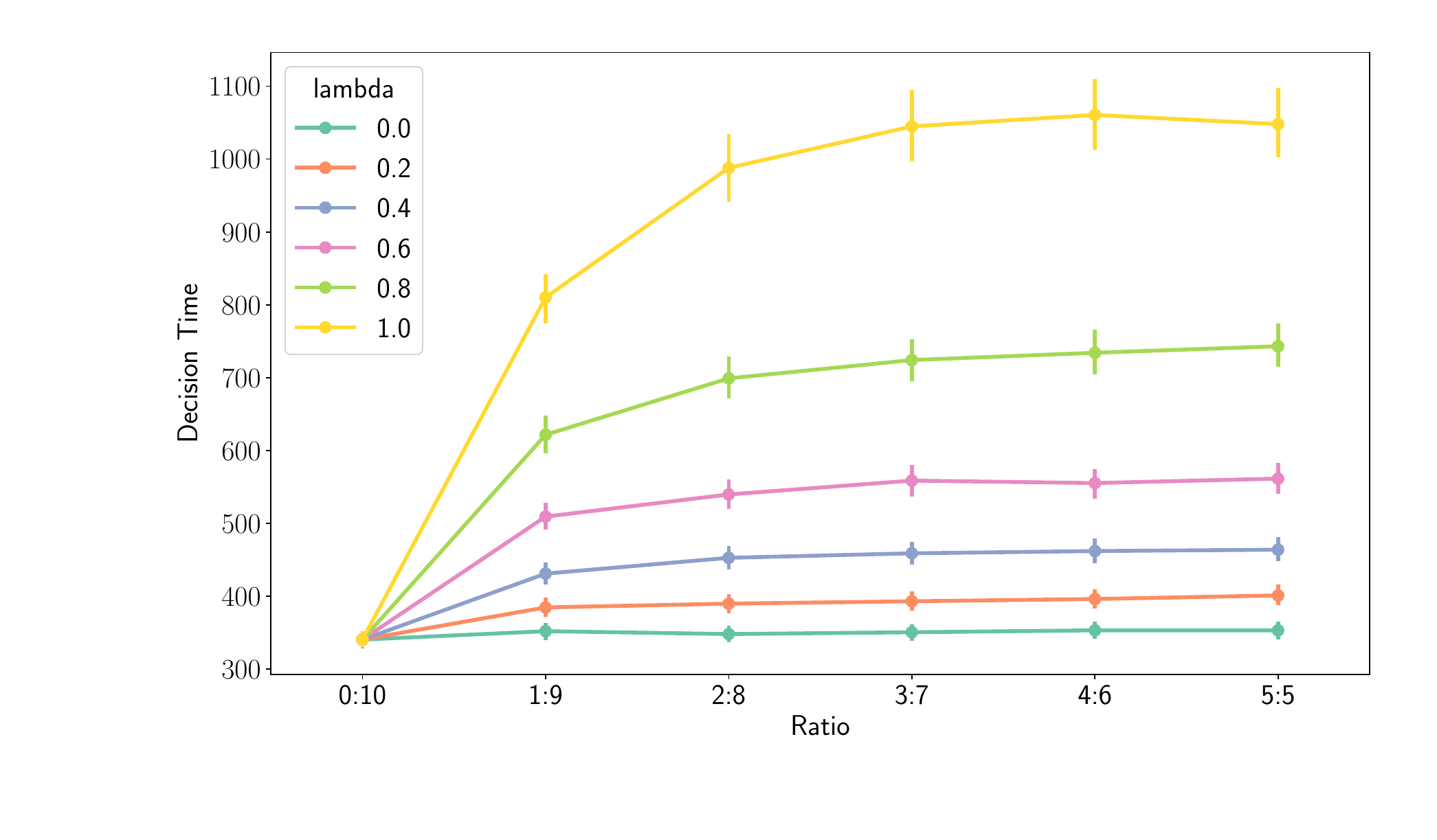}
    \caption{Decision times in the Cambridge Gamble Task. Larger $\lambda$ values (FKL-dominant behavior) yield slower decisions, especially when the red-blue ratio is balanced.}
    \label{fig:cgt_res}
\end{figure*}

In the CGT task, we only show results from the dynamic model since we are interested only in the response times. Indeed, Fig.~\ref{fig:cgt_res} shows decision times for the first stage of the CGT under varying $\lambda$ values (with $\lambda=0$ being a full RKL-based model and $\lambda=1$ a full FKL-based model). As predicted, models dominated by FKL (large $\lambda$) produce slower decisions across all box ratios. This is consistent with empirical findings showing slower responses in autistic participants during CGT trials (\cite{vella2018understanding}). Indeed empirical results tend to show that autistic people make decisions more slowly than neurotypically developing individuals. Interestingly, in the results provided by \cite{vella2018understanding}, decision times from the ASD group showed larger variance than those from the control group. Fig.~\ref{fig:cgt_res} reproduces this trend by revealing that larger $\lambda$-based models, i.e. $\lambda\in [0.5, 1]$, show a larger difference in decision times compared to lower $\lambda$-based models, i.e. $\lambda\in [0, 0.5)$. This means that for a population with individuals that vary in their weighting value $\lambda$, but with large value tendencies, decision times tend to vary more than a population with individuals that have low values.

However, empirical results such as those from \cite{vella2018understanding} also reveal that although ASD groups make decisions more slowly than control groups in all task configurations, i.e., the ratio of blue and red, consistent with an FKL-biased decision process, it also shows that decision times of both groups tend to increase as the ratio becomes more balanced. This suggests that the value of $\lambda$ may not only differ among individuals, but may also depend on the situation.


\section{Discussion}

The present work proposes a unified computational account of decision paralysis that connects phenomenological descriptions of autistic experience with formal models of decision-making. By explicitly separating intention selection (what outcome to pursue) from affordance selection (how to act), and by characterizing inference at each stage in terms of forward- versus reverse-KL divergence, our framework explains how hesitation, freezing, and shutdown can arise even in the absence of learning deficits, noisy value estimates, or downstream executive failure. In both static and dynamic simulations, paralysis emerges when comparison fails to converge at either level of this hierarchy, particularly under inference regimes that preserve multiple plausible alternatives rather than collapsing onto a single dominant option.

\subsection{Intent- and affordance-saturation as distinct failure modes}

A central contribution of this work is the formal distinction between \emph{intent saturation} and \emph{affordance saturation}. Although both manifest behaviourally as slow action or inaction, the model shows that they arise from failures at different stages of the decision-making process. Intent saturation occurs when multiple goals are similarly valued, preventing commitment to a target outcome. Affordance saturation arises when a goal is clear, but multiple actions are equally likely to achieve it. This distinction mirrors first-person autistic reports that differentiate being unable to decide \emph{what} to do from being unable to determine \emph{how} to do it, and it explains why paralysis can occur even when valuation is intact and task structure is fully known.

These failure modes are not specific to autism. Rather, autism functions here as a revealing case in which latent instabilities in a general two-stage decision architecture become more visible. Even in neurotypical cognition, momentary hesitation or freezing can arise when goals or actions are insufficiently differentiated. Autistic experiences magnify these vulnerabilities, clarifying the structural points at which decision-making can fail to converge and motivating a decomposition of choice into intent-level and affordance-level inference.

\subsection{Forward- versus reverse-KL inference and commitment}

Our second key contribution is to link these breakdowns to the divergence principle governing inference. Reverse-KL-dominated processes are mode-seeking and promote sharp commitment, ensuring that a decision is reached even when differences between options are small. Forward-KL-dominated processes, by contrast, are mode-covering and allocate probability mass across multiple plausible alternatives. While such overinclusive representations can support robustness, flexibility, and fidelity to uncertainty, they also render the system vulnerable to non-convergence when many options are similarly valued.

Within the control-as-inference formulation used here, this distinction appears as a geometric constraint: a unimodal neural-state distribution $q(x)$ must approximate a multimodal target distribution $p(x)$ induced by the Boltzmann-optimal intent-action policy. Minimizing reverse KL encourages early collapse of $q(x)$ onto a single mode of $p(x)$, whereas minimizing forward KL penalizes excluding any plausible mode, producing broader, more conservative representations. Decision paralysis arises when this mode-covering tendency prevents the decision variable from entering any decisional basin corresponding to a committed intent or action.

Crucially, FKL-dominated agents in our simulations are not impaired in valuation or choice optimality. Once a decision is reached, their choice distributions closely match those of conventional softmax policies. Paralysis therefore reflects instability on the path to commitment, not suboptimal preferences or noise-driven indecision.

\subsection{Autism, response-time variability, and context dependence}

Cambridge Gamble Task simulations illustrate how this framework accounts for empirical findings in autism. Increasing the weight of forward-KL inference produces both slower mean decision times and greater variability, consistent with reports of prolonged and more variable response times in autistic participants \citep{vella2018understanding}. Notably, this variability emerges without invoking heterogeneous noise, attentional lapses, or executive dysfunction. It follows directly from the geometry of inference under multimodal value structure.

At the same time, both autistic and neurotypical agents in our model show increased decision times as evidence becomes more ambiguous, suggesting that the balance between forward- and reverse-KL influences is unlikely to be a fixed trait parameter. Instead, the mixture parameter $\lambda$ may be context dependent, modulated by factors such as uncertainty, cognitive load, or stress. Autism may be characterized by a higher baseline tendency toward FKL-dominated inference or by greater sensitivity of $\lambda$ to environmental ambiguity, leading to disproportionate delays in open-ended or multi-option contexts.

\subsection{Relation to Bayesian and predictive-processing accounts}

The present account complements existing Bayesian and predictive-processing theories of autism, which emphasize atypical prior precision, volatility estimation, or prediction-error weighting \citep{lawson2017predictive,lieder2019slow}. Empirical support for these accounts is mixed and highly task dependent, with some studies reporting atypical updating and others finding largely intact predictive processing under certain conditions \citep{pesthy2023intact,shi2025predictive}. The forward-reverse KL distinction introduces an orthogonal dimension that does not compete with these proposals, but instead characterizes the \emph{optimization geometry} of inference.

Differences in mode-covering versus mode-seeking tendencies can help explain why atypical updating appears in some paradigms, but not others. An FKL-dominant regime naturally produces conservative belief dynamics by penalizing exclusion of previously plausible hypotheses, yielding slower reallocation of probability mass when contingencies change. Although our simulations focus on decision variables rather than perceptual beliefs, the same KL geometry could in principle operate across hierarchical levels of inference.

\subsection{Relation to existing computational decision models}

Several established computational traditions address aspects of decision instability, but none explicitly capture the full structure of the breakdowns formalized here. Classical accumulator frameworks, such as drift-diffusion models, explain slowed or unstable choices when evidence or subjective values are similar, but assume that intentions and action options are already well defined \citep{milosavljevic2010drift, fontanesi2019reinforcement, bogacz2006physics}. These models also do not account for why such instabilities appear more pronounced in autism. Neuroeconomic and reinforcement-learning approaches similarly formalize choice as competition among alternative policies or value estimates (typically via softmax selection), while leaving unspecified those processes by which intentions and action channels are constructed \citep{sutton2018reinforcement, rangel2008framework}. As a result, they invariably generate a choice, and cannot represent paralysis. Predictive-processing accounts of autism highlight atypical estimates of sensory precision and diminished priors \citep{pellicano2012world, lawson2014aberrant}, offering a potential source of overload, but unable to explain the specific phenomenology of inertia or shutdown. Meanwhile, phenomenological and qualitative studies \citep{dekker1999our, buckle2021no, welch2021Inertia, phung2021BIMS, ayaya2022asd, rapaport2024live} document pervasive difficulties with action initiation and overload-induced collapse, and affordance-based accounts note differences in how possibilities for action are perceived in autism \citep{hellendoorn2014understanding}. Yet the literature remains fragmented, leaving unresolved how intention formation, affordance extraction, and action commitment interact to produce paralysis in both typical and autistic cognition.
By explicitly separating intention formation, affordance extraction, and action commitment, the present framework integrates these otherwise disconnected insights into a single computational account of paralysis.

\subsection{Implications beyond choice}

Although the present paper focuses on decision paralysis, the core mathematical property of forward-KL minimization (its resistance to premature collapse) offers a unifying perspective on several cognitive characteristics associated with autism. Broad, high-fidelity representations are consistent with enhanced attention to detail and reduced susceptibility to certain visual illusions \citep{pellicano2012world,chrysaitis2023tenyears}, while conservative inference may contribute to inflexibility in task switching or reduced spontaneity in generating actions under open-ended demands. Importantly, this framework does not imply a global deficit in creativity or flexibility. Rather, it predicts selective constraints in contexts that require rapid abstraction, switching, or commitment under ambiguity, consistent with the heterogeneous empirical literature \citep{ayaya2022asd,cancer2024creative}.

\subsection{Limitations and future directions}

Several limitations warrant consideration. First, tasks examined here are intentionally simple and static, allowing us to isolate decision dynamics from learning and environmental uncertainty. Extending the model to sequential, partially observable, or socially embedded settings will be essential to capture real-world paralysis. Second, we did not fit the KL mixture parameter $\lambda$ to participant-level data. Links to autistic cognition should therefore be interpreted as mechanistic hypotheses rather than quantitative claims. Third, although the model is compatible with opponent-inhibition and drift-diffusion architectures, we have not specified a neural implementation. Relating KL geometry to neurophysiological mechanisms, such as excitation-inhibition balance, neuromodulation, or fronto-striatal dynamics, remains an important direction for future work.

\subsection{Conclusion}

By formalizing decision paralysis as a failure of convergence in a hierarchical decision process governed by the balance between forward- and reverse-KL inference, this work provides a unified account of hesitation, inertia, and shutdown across typical and autistic cognition. Autistic experiences are not treated as exceptions, but as magnified expressions of general computational trade-offs inherent in human decision-making. In doing so, the present framework offers both a conceptual bridge between lived experience and formal theory, and a foundation for future empirical tests of how inference geometry shapes ability to commit to action.

\section{Methods}\label{sec:methods}

In this section we present the conceptual structure of our model, while full mathematical derivations and technical details are provided in Supplementary Information~\ref{apdx:methods}.

\subsection{The Kullback--Leibler divergence}

We use the Kullback--Leibler (KL) divergence to formalize how an agent compares two probability distributions. Given distributions $q(x)$ and $p(x)$, the KL divergence measures the expected evidence each sample provides against one distribution if the other generated the data. Crucially, the KL divergence is \emph{asymmetric}:
\begin{align}
D_{\mathrm{KL}}(q \mid\mid p) \neq D_{\mathrm{KL}}(p \mid\mid q).
\end{align}

This asymmetry leads to two qualitatively different behaviors:
\begin{itemize}
    \item \emph{Reverse KL}, $D_{\mathrm{KL}}(q \mid\mid p)$, is \emph{mode-seeking}: the approximation tends to collapse onto the highest-probability mode of $p$.
    \item \emph{Forward KL}, $D_{\mathrm{KL}}(p \mid\mid q)$, is \emph{mode-covering}: the approximation spreads to cover all regions where $p$ has support.
\end{itemize}

This distinction is central to our model, where different KL objectives produce different forms of decision behavior.

\subsection{Control as Inference: Decision-Making as Probabilistic Inference}

We adopt the Control-as-Inference (CaI) framework (see Section~\ref{apdx:methods_cai} and/or \cite{levine2018reinforcement}), which formulates decision-making as approximating a target distribution over optimal trajectories. A policy $\pi$ induces a distribution over trajectories $\tau$, while the reward function defines a target distribution $p^{*}(\tau)$ in which high-reward trajectories are exponentially preferred.

The optimal policy is defined as the minimizer of the reverse KL divergence:
\begin{align}
\pi^{*} = \argmin{\pi}{D_{\mathrm{KL}}\!\bigl( p^{\pi}(\tau) \mid\mid p^{*}(\tau) \bigr)}.
\end{align}

This minimization yields the familiar Boltzmann (softmax) policy:
\begin{align}
\pi(a \mid s) \propto \exp\!\left( \frac{1}{\beta} Q(s,a) \right),
\end{align}
which we use to specify the probabilistic preferences among actions.

\subsection{Model Overview}

Our model links this probabilistic decision framework to a neural decision variable $x$ that must eventually commit to a single option. The construction follows three conceptual steps (mathematical details for each concept are given in Supplementary Information~\ref{apdx:methods_desc}).

\paragraph{Neural encoding of options}

Each action--outcome pair $k$ is associated with a Gaussian region in neural state space:
\begin{align}
p(x \mid k) = \mathcal{N}(x \mid \mu_k, \Sigma_k).
\end{align}

\paragraph{Target neural-state distribution}

The Boltzmann policy assigns probability $\pi_k$ to each option, producing a mixture distribution over neural states:
\begin{align}
p(x) = \sum_{k} \pi_k\, p(x \mid k).
\end{align}

\paragraph{Unimodal neural representation}

Because the agent must ultimately commit to one choice, the neural state is modeled as a unimodal distribution $q(x)$ that approximates $p(x)$. We define this approximation as the minimizer of a mixture of forward and reverse KL divergences:
\begin{align}
q^{*} = \argmin{q}{\Bigl[ (1-\lambda)\,D_{\mathrm{KL}}(q \mid\mid p) \;+\; \lambda\,D_{\mathrm{KL}}(p \mid\mid q) \Bigr]}
\end{align}

Small values of $\lambda$ (reverse KL dominance) produce sharp, decisive choices, while large values of $\lambda$ (forward KL dominance) may produce diffuse or unstable states, particularly when options are similarly valued. We hypothesize that autistic individuals may operate with effectively larger $\lambda$, making them more prone to indecision under ambiguity.

\subsection{Dynamic Drift--Diffusion Formulation}

To describe the temporal evolution of decisions, we transform the KL minimization problem into a drift--diffusion process for the neural variable $x_t$. Interpreting the KL objective as defining a Wasserstein gradient flow (\cite{santambrogio2017euclidean}) yields a stochastic differential equation of the form:
\begin{align}
dx_t = \mu(x_t)\,dt + \sigma\, dW_t,
\end{align}
where the drift $\mu(x_t)$ depends on the chosen KL objective.

Specifically, Reverse-KL-dominated dynamics produce strong drifts toward a single mode of $p(x)$, yielding fast and decisive trajectories. In contrast, Forward-KL-dominated dynamics produce weak drifts in low-probability regions of $p$, yielding slow, indecisive trajectories. Complete drift expressions for the forward, reverse, and mixed $\lambda$-KL objectives are provided in Section~\ref{apdx:methods_desc}.

\subsection{Opponent-Inhibition Decision Dynamics}

To connect the probabilistic model with cortical mechanisms, we interpret each mixture component $p_k(x)$ as a population of neurons selective for option $k$. This allows us to incorporate \emph{self-excitation} (pulling activity toward the chosen option) and \emph{cross-inhibition} (suppressing alternatives), capturing the opponent-inhibition motif found in the literature (\cite{kuan2024synaptic}). The resulting dynamics combine an excitatory drift toward $p(x)$ with an inhibitory drift emphasizing alternatives not chosen.

The final stochastic differential equation governing the neural decision dynamics is:
\begin{align}
dx_t &= \mu_t\, dt + \sqrt{2}\, dW_t, \\
\mu_t &= \sum_k \bigl( \mu_{t,k}^{E} - \mu_{t,k}^{I} \bigr),
\end{align}
where $\mu_{t,k}^{E}$ and $\mu_{t,k}^{I}$ are the excitatory and inhibitory drifts associated with option $k$, i.e. respectively, the drift toward option $k$ from self-excitation and the drift away from option $k$ due to inhibition by competing options.
Full mathematical specification is provided in the Supplementary Information (Section~\ref{apdx:methods_desc}).

\subsection{Summary}

\begin{figure*}[b]
    \centering
    \includegraphics[keepaspectratio=true,width=\linewidth]{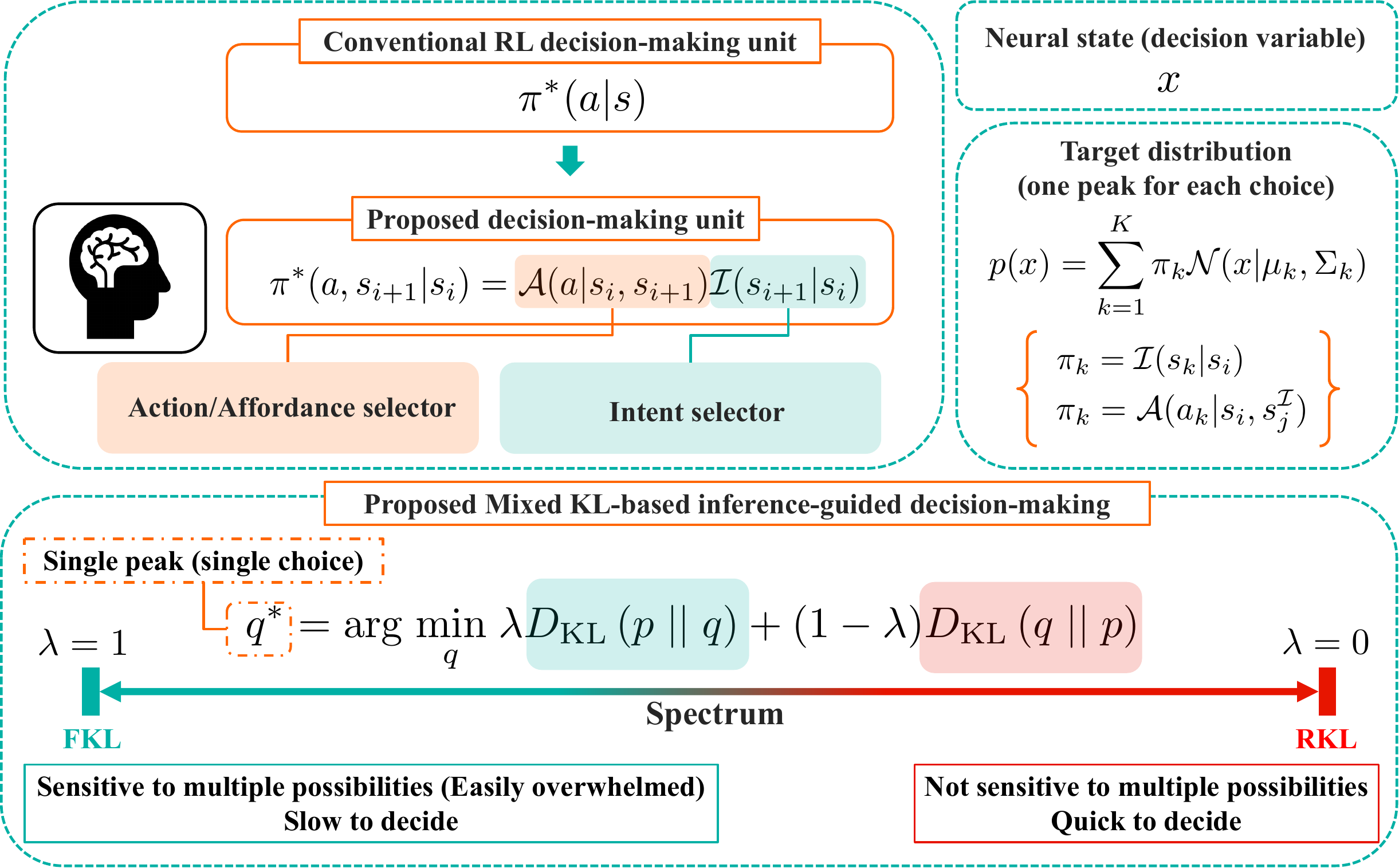}
    \caption{Overview of the proposed computational model of intent and affordance saturation.}
    \label{fig:graphical_abstract}
\end{figure*}

In summary:
\begin{itemize}
    \item Decision-making is formulated as probabilistic inference over trajectories.
    \item Each option corresponds to a neural attractor; the brain forms a unimodal approximation $q(x)$.
    \item The balance between forward and reverse KL determines whether the system is decisive or indecisive, i.e. prone to slow decision or decision paralysis, and is controlled by a parameter $\lambda$ (Fig.~\ref{fig:graphical_abstract}).
    \item Adding opponent-inhibition dynamics yields a biologically grounded KL-based drift--diffusion model.
\end{itemize}

This framework provides a computational account of individual differences in decision-making behavior, arising from different effective balances of forward and reverse KL contributions.

\section{Data availability}

No underlying data are available for this article, since no datasets were generated or analysed during this study.

\section{Code availability}

All code is available via GitHub at \url{https://github.com/Mahoumaru/kl_based_asd}

\section{Acknowledgements}

We thank Sho Yagishita, Satsuki Ayaya, Shin-ichiro Kumagaya, and Kiyoto Kasai for helpful discussions.

This work was supported by JSPS KAKENHI Grant Numbers JP21H05172 and JP21H05171.

\section{Author contributions}

\begin{itemize}
\item[] \textbf{Wendyam Eric Lionel Ilboudo} contributed to Conceptualization, Methodology, Software, Investigation, Validation, Visualization and Writing - Original Draft
\item[] \textbf{Saori C Tanaka} contributed to Resources, Supervision, Conceptualization, Validation, Funding acquisition and Writing - Review \& Editing
\end{itemize}

\section{Competing interests}

The authors declare no competing interests.

\bibliography{sn-bibliography}

\newpage

\begin{appendices}

\section{Methods: Supplementary Information and detailed equations}\label{apdx:methods}

\subsection{The Kullback-Leibler divergence}\label{apdx:methods_kl}

 Given two distributions $q(x)$ and $p(x)$ over a random variable $x$, the Kullback-Leibler (KL) divergence between $q$ and $p$, denoted as $D_{\mathrm{KL}}(q \mid\mid  p)$, is a statistical distance measuring how different $q$ is from $p$. Specifically, the KL divergence is the average weight of evidence (or average log-likelihood ratio $\log \frac{q(x)}{p(x)}$), i.e., the expected per-sample evidence against $p$ under the assumption that  samples are truly sampled from $q$.

In other words, it is the typical amount of evidence we expect each new observation $x$ to give us against an alternative distribution $p$ if the true distribution from which the samples are generated is $q$. If $q$ and $p$ are identical, then they are indistinguishable, the weight of evidence averages to zero, and the KL divergence vanishes.
 
 Mathematically, the KL divergence is defined as:
\begin{align}
D_{\mathrm{KL}}\!\bigl( q \,\mid\mid \, p \bigr)
&= \lim_{N \to \infty} \frac{1}{N} \sum_{k=1}^{N} \log \frac{q(x_k)}{p(x_k)}
= \sum_x q(x)\,\log \frac{q(x)}{p(x)} \\
&= \mathbb{E}_{x \sim q(x)}\bigl[ \log \frac{q(x)}{p(x)} \bigr]
\end{align}
In the continuous case, $\sum_x$ is replaced by an integral over $dx$.

Key to our discussion, the KL divergence is an asymmetric divergence measure. Indeed,
\[
D_{\mathrm{KL}}\bigl(q \mid\mid  p \bigr) \neq D_{\mathrm{KL}}\bigl(p \mid\mid  q \bigr).
\]
This is because $D_{\mathrm{KL}}\bigl(q \mid\mid  p \bigr)$ and $D_{\mathrm{KL}}\bigl(p \mid\mid  q \bigr)$ measure two different things.
Specifically,
\begin{itemize}
  \item $D_{\mathrm{KL}}(q\mid\mid p)$ is the expected weight of evidence \emph{against $p$}, if data is generated by $q$;
  \item $D_{\mathrm{KL}}(p\mid\mid q)$ is the expected weight of evidence \emph{against $q$}, if data is generated by $p$.
\end{itemize}
Thus, the choice of divergence for distinguishing between two distributions depends on which distribution is assumed to be the true data-generating process.

The underlying assumption and the asymmetry have important consequences when the KL divergence is employed to approximate unknown distributions, i.e.\ to find a distribution $q$ that closely matches a target distribution $p$ in terms of its behavior on the sample space. In particular:

\begin{itemize}
  \item The \emph{Forward KL} optimization, i.e. $\min_q D_{\mathrm{KL}}(p \,\mid\mid \, q)$, penalizes heavily when $q(x)$ assigns very low probability to regions where $p(x)$ has support. As a result, minimizing the forward KL encourages $q$ to cover all modes of $p$, even if this means spreading probability mass more broadly. This behavior is often described as \emph{mode-covering} or \emph{mean-seeking}.
  \item The \emph{Reverse KL} optimization, i.e. $\min_q D_{\mathrm{KL}}(q \,\mid\mid \, p)$, penalizes heavily when $q(x)$ puts mass in regions where $p(x)$ is small. Thus, minimizing the reverse KL encourages $q$ to concentrate on a single high-probability mode of $p$, possibly ignoring other modes. This is referred to as \emph{mode-seeking}.
\end{itemize}

These contrasting behaviors explain why forward and reverse KL are suited to different applications. For example, variational inference methods often minimize the reverse KL (leading to mode-seeking approximations), while expectation propagation minimizes the forward KL (leading to mode-covering approximations). Below we show how the softmax policy typically employed in RL relates to the reverse KL.

\subsection{Control-as-Inference or Value-based Decision-Making as Inference}\label{apdx:methods_cai}

Control as Inference (CaI - \cite{levine2018reinforcement}) is a theoretical framework in which optimal control and reinforcement learning (RL) are formulated as reasoning on probabilistic graphical models. Specifically, CaI formulates the RL problem as trying to find a policy (or decision-making strategy) that generates a distribution over trajectories which approximates the distribution over optimal trajectories defined by the reward function.

 Specifically, let $\tau = \{ s_1, a_1, s_2, a_2, \cdots, s_T, a_T, s_{T+1} \}$ be a trajectory, i.e. sequence of states and actions, then we can write the probability of $\tau$ as:
\begin{align*}
 p^{\pi}(\tau) = p(s_1, a_1, s_2, a_2, \cdots, s_T, a_T, s_{T+1}) = p(s_1)\prod\limits_{t=1}^{T} p(s_{t+1} | s_t, a_t) \pi(a_t | s_t)
\end{align*}
where $p(s_{t+1} | s_t, a_t)$ is the transition probability that defines dynamics of the environment and $\pi(a_t | s_t)$ is some given decision-making rule which selects the action based on the current state.

The control problem is in practice concerned with finding an optimal trajectory with respect to the cumulative sum of rewards. This reward signal therefore defines a distribution over optimal trajectories given by:
\begin{align}
 p^{*}(\tau) &\propto p(s_1)\prod\limits_{t=1}^{T} p(s_{t+1} | s_t, a_t) \exp\left( \frac{1}{\beta} r(s_t, a_t) \right) \nonumber \\
 &= \left[ p(s_1)\prod\limits_{t=1}^{T} p(s_{t+1} | s_t, a_t) \right] \exp\left( \frac{1}{\beta} \sum\limits_{t=1}^{T} r(s_t, a_t) \right)
 \label{eq:cai_target}
\end{align}
 Intuitively, we can see that the larger the sum of rewards is, the more likely this trajectory becomes. Hence, this distribution naturally samples trajectories that are optimal with respect to the reward function.
 
 The control as inference approach to RL is concerned with finding a policy $\pi$ which induces a distribution $p^{\pi}(\tau)$ that produces sampled trajectories indistinguishable from samples generated by $p^{*}(\tau)$. For this purpose, CaI minimizes the reverse KL divergence:
\begin{align}
\pi^{*} = \argmin{\pi}{D_{\mathrm{KL}}(p^{\pi}(\tau) \,\mid\mid \, p^{*}(\tau))}
\end{align}
 The optimal policy $\pi^{*}$ that minimizes this KL divergence objective can be computed recursively and is given by the Boltzmann distribution:
\begin{align}
\pi^{*}(a_t | s_t) = \exp\left( \frac{1}{\beta} Q(s_t, a_t) - \frac{1}{\beta}V(s_t) \right) = \frac{\exp\left( \frac{1}{\beta}Q(s_t, a_t) \right)}{\sum_a \exp\left( \frac{1}{\beta}Q(s_t, a) \right)} \\
\mathrm{where}\;\; Q(s_t, a_t) = r(s_t, a_t) + \mathbb{E}_{p(s_{t+1} | s_t, a_t)}\left[ V(s_{t+1}) \right]
\end{align}
$V(s_t) = \beta \log \sum_a \exp\left( \frac{1}{\beta}Q(s_t, a) \right)$ is the normalization constant, independent of the action. A quick look back at equation \ref{eq:softmax} clearly shows that this is equivalent to the softmax policy typically used in computational neuroscience decision-making models.
Hence, the softmax policy originates from the reverse KL divergence. Similarly, the maximum operator for deterministic action selection can also be shown to be related to the RKL.

\subsection{Model Description and Implementation}\label{apdx:methods_desc}

We construct our computational model of decision-making on top of the control-as-inference (CaI) formulation. In this framework, action selection is cast as approximate inference over trajectories: the agent seeks to approximate a target distribution $p^{*}(\tau)$ as defined in equation~\eqref{eq:cai_target}. This distribution is generally multimodal, since multiple optimal or near-optimal trajectories may exist.

An agent that must commit to a single choice cannot, however, represent multiple incompatible trajectories simultaneously. Consequently, the variational distribution $q(\tau)$ used to approximate $p^{*}(\tau)$ must be unimodal. This observation rules out using the Boltzmann policy as a direct decision rule: as the exact reverse-KL minimizer, it can remain multimodal in order to match $p^{*}(\tau)$. Nonetheless, the Boltzmann policy remains useful for defining a target distribution over actions (or trajectories) that the agent tries to approximate as demonstrated in the next section.

Throughout, we assume a finite Markov decision process (MDP) with finitely many states and actions, as is typical in computational neuroscience tasks. To bridge this discrete decision space with continuous neural activity, we consider a continuous variable $x$ representing the neural state of the agent, in alignment with drift-diffusion models.

\subsubsection{Static (time-independent) model}

\paragraph{Intent and affordance selectors}

Given the current state $s$, the agent must decide both which outcome $s'$ it intends to reach next, and which action $a$ will realize that outcome. To capture this two-step process, we replace the standard action-only policy $\pi(a\mid s)$ with a joint intent--action policy
\begin{align}
\pi(s', a \mid s) = \mathcal{I}(s' \mid s)\,\mathcal{A}(a \mid s, s'),
\end{align}
where $\mathcal{I}(s'\mid s)$ is the \emph{intent selector} (distribution over desired next states) and $\mathcal{A}(a\mid s,s')$ is the \emph{affordance (action) selector} that chooses an action conditioned on both the current state and the intended next state.

Under this joint policy, the induced trajectory distribution is
\begin{align}
p^{\pi}(\tau)
 &= p(s_1)\prod_{t=1}^{T} \pi(s_{t+1}, a_t \mid s_t) \\
 &= p(s_1)\prod_{t=1}^{T} \mathcal{I}(s_{t+1} \mid s_t)\,\mathcal{A}(a_t \mid s_t, s_{t+1}).
\end{align}
Using the environment transition dynamics $p(s_{t+1}\mid s_t,a_t)$, we may express $\mathcal{I}$ and $\mathcal{A}$ in terms of the original action policy:
\begin{align}
\mathcal{I}(s_{t+1}\mid s_t)
  &= \int \pi(s_{t+1}, a_t \mid s_t)\, da_t
   = \int p(s_{t+1}\mid s_t, a_t)\,\pi(a_t \mid s_t)\, da_t, \\
\mathcal{A}(a_t \mid s_t, s_{t+1})
  &= \frac{p(s_{t+1}\mid s_t, a_t)\,\pi(a_t\mid s_t)}{\mathcal{I}(s_{t+1}\mid s_t)}.
\end{align}
The optimal intent and action selectors $(\mathcal{I}^{*}, \mathcal{A}^{*})$ are obtained by substituting the Boltzmann-optimal policy $\pi^{*}(a\mid s)$ into the above expressions.

\paragraph{Neural-state variable $x$}

Suppose the agent selects an intent--action pair $z_k = (s^{(k)}_{t+1}, a^{(k)}_{t})$, where $k \in \{1,\dots,K\}$ with $K = N_s N_a$, $N_s$ and $N_a$ being the numbers of states and actions, respectively. We associate with each choice $z_k$ a region of neural activity through the conditional distribution
\begin{align}
p(x \mid z_k) = \mathcal{N}(x \mid \mu_k, \Sigma_k),
\end{align}
where $x \in \mathbb{R}^{K-1}$ and $(\mu_k,\Sigma_k)$ characterize the firing-rate region corresponding to choice $k$. The mean $\mu_k$ is the most likely neural configuration when choice $k$ dominates, and $\Sigma_k$ captures biological noise and variability. Together, $(\mu_k,\Sigma_k)$ define a delimited region in which we are likely to find $x$ if $z_k$ is selected. Specifically, we say that the agent selects choice $k$ if the neural state $x$ lies within the Mahalanobis decision region associated with $p_k(x)$. Concretely, the decision is
\emph{choice $k$} whenever
\begin{align}
(x - \mu_k)^{\top} \Sigma_k^{-1} (x - \mu_k) \;\le\; \delta_k,
\end{align}
where $\delta_k$ is a prescribed threshold defining the boundary of the decision
region for option $k$. In practice we set $\Sigma_k$ as the identity matrix and $\delta_k = 1$, for all choice $k$.

Using these components, the target distribution over neural states at decision time is
\begin{align}
p(x) &= \sum_{k=1}^{K} \pi_k\, \mathcal{N}(x \mid \mu_k, \Sigma_k), \\
\pi_k &= \pi^{*}(s'=s^{(k)}, a=a^{(k)}),
\end{align}
and the actual neural state $x$ is drawn from a unimodal variational approximation $q(x)$ found by solving
\begin{align}
q^{*}(x) = \argmin{q}{\Bigl[ (1-\lambda) D_{\mathrm{KL}}(q(x)\,\mid\mid \,p(x)) + \lambda D_{\mathrm{KL}}(p(x)\,\mid\mid \,q(x)) \Bigr]}.
\end{align}

For $\lambda = 0$ (reverse KL), $q^{*}$ collapses onto one of the mixture components: specifically, $q^{*}(x) = p(x \mid z_k)$ with probability $\pi_k$. The agent thus commits to a categorical decision, selecting one option with probability proportional to its Boltzmann weight.

\begin{figure}[h]
    \centering
    \includegraphics[keepaspectratio=true,width=\linewidth]{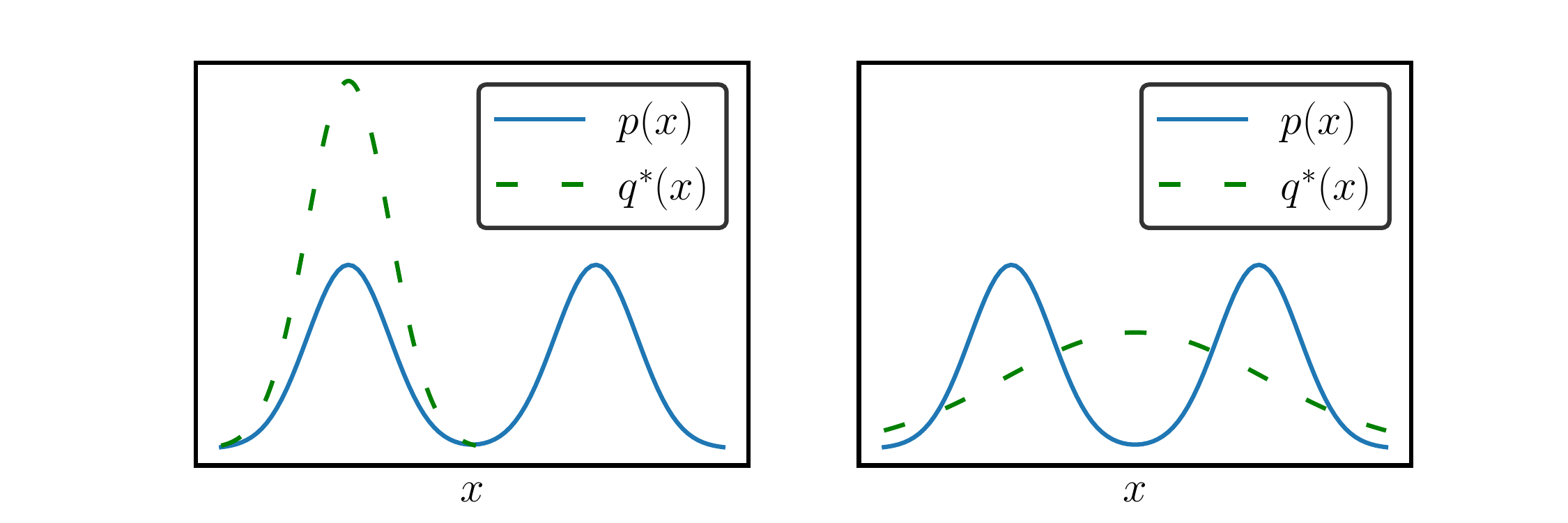}
    \caption{Visualization of how the Forward KL divergence ($\lambda = 1$) can result in decision paralysis when facing equally valued options in the case of two options, e.g. single state/intent with two possible actions. Left: $\lambda = 0$ (RKL); Right: $\lambda = 1$ (FKL).}
    \label{fig:2choice_case}
\end{figure}

For $\lambda = 1$ (forward KL), $q^{*}$ is a broad distribution whose mean $\mu^{*}$ typically lies between the choice-specific regions and whose covariance $\Sigma^{*}$ is large enough to cover multiple modes roughly in proportion to $\pi_k$. When options are similarly valued (i.e., the $\pi_k$ are comparable), the agent can be driven into regions of $x$ that do not correspond to any valid choice region, leading to \emph{decision paralysis} (Fig.~\ref{fig:2choice_case}). In this paper we hypothesize that autistic individuals may exhibit effectively larger $\lambda$, and are therefore more prone to such paralysis than neurotypical individuals.

\subsubsection{Dynamic (time-dependent) drift-diffusion formulation}

The static analysis above characterizes the distribution of $x$ at the moment of choice, but not its temporal evolution. Empirically, neural decision variables evolve over time from an undecided region toward decision-selective attractors. To capture these dynamics, we derive a drift-diffusion-like process for $x_t$ based on Wasserstein gradient flows (\cite{santambrogio2017euclidean}), which naturally lead to stochastic differential equations (SDEs).

\paragraph{From KL minimization to a probability-flow equation}

Let $x_t \sim q_t(x)$ for all $t$, and define $q_t$ as the minimizer of a single implicit Euler (JKO) step:
\begin{align}
q_t = \argmin{q}{\Bigl[ \mathcal{F}(q,p) + \frac{\mathcal{W}_2^2(q,q_{t-1})}{2\tau} \Bigr],}
\end{align}
where $\mathcal{W}_2$ is the 2-Wasserstein distance, $\tau$ is a time step, and $\mathcal{F}$ is either $D_{\mathrm{KL}}(q\mid\mid p)$, $D_{\mathrm{KL}}(p\mid\mid q)$, or a convex combination thereof. Taking $\tau \to 0$ yields the Wasserstein gradient flow
\begin{align}
\frac{\partial q_t}{\partial t} = -\nabla_x \cdot \bigl( q_t(x)\, v_t(x) \bigr),
\label{eq:proba_flow_rewrite}
\end{align}
where the velocity field $v_t$ is given by the gradient of the functional derivative of $\mathcal{F}$, i.e.:
\begin{align}
v_t(x) = -\nabla_x \left[\frac{\delta \mathcal{F}(q,p)}{\delta q}\right].
\label{eq:velocity_rewrite}
\end{align}

From the theory of SDEs, the process
\begin{align*}
dx_t = \mu(x_t)\,dt + \sigma(x_t)\,dW_t
\end{align*}
induces the Fokker--Planck equation
\begin{align}
\frac{\partial q_t}{\partial t}
  = -\nabla_x \cdot \Bigl( q_t(x) \bigl[ \mu(x) - D(x)\nabla_x \ln q_t(x) - \nabla_x D(x) \bigr] \Bigr),
\end{align}
where $D(x) = \tfrac{1}{2}\sigma(x)\sigma(x)^{T}$ is the diffusion (mobility) tensor. Identifying this equation with~\eqref{eq:proba_flow_rewrite} gives
\begin{align}
\mu(x_t) &= v_t(x_t) + D(x_t)\,\nabla_x \ln q_t(x_t) + \nabla_x D(x_t).
\label{eq:drift_general_rewrite}
\end{align}

Crucially, the gradient-flow PDE fixes $v_t$ through $\mathcal{F}$, but it does not uniquely determine $(\mu, D)$: any smooth symmetric positive definite tensor $D(x)$ is admissible, with $\mu$ then constrained by~\eqref{eq:drift_general_rewrite}. Thus $D$ is a modeling choice that sets the effective geometry or mobility of the diffusion. In our simulations we adopt the simplest choice $D(x) = 1$, leading to
\begin{align}
dx_t = \bigl[ v_t(x_t) + \nabla_x \ln q_t(x_t) \bigr] dt + \sqrt{2}\,dW_t.
\end{align}

\paragraph{Forward vs.\ reverse KL dynamics}

Let $f(r) = \lambda r\ln r - (1-\lambda)\ln r$, and define the objective function as an f-divergence:
\begin{align}
\mathcal{F}_\lambda(q,p)
  = (1-\lambda)D_{\mathrm{KL}}(q\mid\mid p) + \lambda D_{\mathrm{KL}}(p\mid\mid q)
  = \mathbb{E}_{x\sim q} \left[ f\!\left(\frac{p(x)}{q(x)}\right) \right].
\end{align}
Then
\begin{align}
v_t(x)
 &= \left(\frac{p(x)}{q_t(x)}\right)^{2} f''\!\left(\frac{p(x)}{q_t(x)}\right)
    \bigl( \nabla_x \ln p(x) - \nabla_x \ln q_t(x) \bigr),
\end{align}
with
\begin{align}
f''(r) = \frac{\lambda}{r} + \frac{1-\lambda}{r^2}.
\end{align}
This yields
\begin{align}
v_t(x) =
\begin{cases}
\nabla_x \ln p(x) - \nabla_x \ln q_t(x), & \lambda=0\ \text{(RKL)}, \\[3pt]
\frac{p(x)}{q_t(x)} \bigl[ \nabla_x \ln p(x) - \nabla_x \ln q_t(x) \bigr], & \lambda=1\ \text{(FKL)}, \\[3pt]
\bigl[(1-\lambda) + \lambda \tfrac{p(x)}{q_t(x)}\bigr]
  \bigl[ \nabla_x \ln p(x) - \nabla_x \ln q_t(x) \bigr], & 0<\lambda<1.
\end{cases}
\end{align}

For the reverse KL ($\lambda=0$), $v_t(x)$ is the gradient (with respect to $x$) of the log-likelihood ratio between $p$ and $q_t$. It therefore drives the process toward modes of $p(x)$ and away from modes of $q_t(x)$, maximizing a kind of ``weight of evidence'' for the target. In contrast, for the forward KL ($\lambda=1$), the velocity is modulated by the likelihood ratio $p(x)/q_t(x)$. When $x$ lies in a low-probability region under $p$, this ratio is small, resulting in a weak drift and a slow moving process; precisely the situation we would expect in an extended indecision regime.

In our simulations we use the stationary FKL minimizer $q^{*}$, i.e.\ $q_t = q^{*} = \argmin{q}{ D_{\mathrm{KL}}(p\mid\mid q)}$, and obtain the $\lambda$-KL drift
\begin{align}
dx_t
 &= \Bigl[ \bigl(1-\lambda + \lambda \tfrac{p(x_t)}{q^{*}(x_t)}\bigr)\,
     \nabla_x \ln p(x_t)
     + \lambda\bigl(1-\tfrac{p(x_t)}{q^{*}(x_t)}\bigr)\,
       \nabla_x \ln q^{*}(x_t)
     \Bigr] dt
     + \sqrt{2}\,dW_t.
\end{align}
Note that with the choice of $q^{*}$, the $\lambda$-KL velocity is an interpolation between the fast RKL's Langevin drift (equal to $\nabla_x \ln p(x_t)$ when $\lambda=0$) in which $q^{*}$ has no effect and the slow FKL velocity.

\paragraph{Opponent inhibition model}

Since $p(x)$ is a mixture of Gaussians,
\begin{align}
p(x) = \sum_{k=1}^{K} \pi_k\, p_k(x), \qquad p_k(x) = \mathcal{N}(x\mid\mu_k,\Sigma_k),
\end{align}
its log-gradient can be written
\begin{align}
\nabla_{x} \ln p(x)
 = \sum_{k=1}^{K} \frac{\pi_k}{p(x)}\,\nabla_{x} p_k(x).
\end{align}
This expression naturally corresponds to competition among choice-selective excitatory populations: each term represents the ``pull'' from population $k$ with strength proportional to $\pi_k$ and modulated by the current state $x$ (indeed, $\nabla_{x} p_k$ points towards region $k$).

\begin{figure}[h]
    \centering
    \includegraphics[keepaspectratio=true,width=\linewidth]{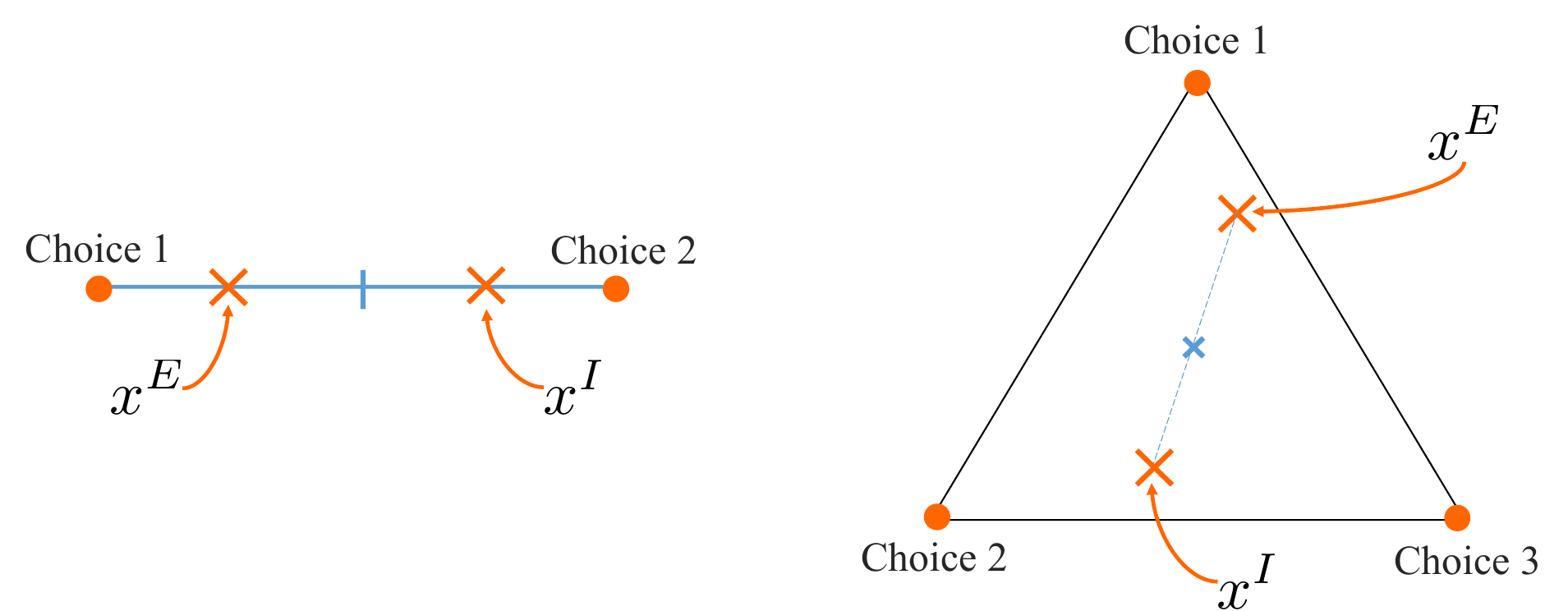}
    \caption{Mirror relation between the excitatory and inhibitory processes. When $x^{E}$ is close to choice 1, this implies that the choice-selective neuronal population for choice 1 is highly excited. Hence, the alternative choices are also strongly inhibited (represented by $x^{I}=-x^{E}$ being closer to the alternative choices).}
    \label{fig:xE_vs_xI}
\end{figure}

This appears as a single part of an opponent-inhibition model. Indeed, opponent-inhibition (OI) models posit that each choice-selective population both excites itself and inhibits the others~(\cite{kuan2024synaptic}). In particular, for each choice $k$:
\begin{itemize}
  \item excitatory neurons promote their own option (self-excitation), stabilizing that choice;
  \item inhibitory neurons suppress competing populations (cross-inhibition), disfavoring alternative options.
\end{itemize}
To capture this within our framework, we introduce excitatory and inhibitory activity variables $x^{E}$ and $x^{I}$ with a mirror relation $x^{I} = -x^{E} = -x$ (Fig.~\ref{fig:xE_vs_xI}). When $x^{E}$ lies near the region for choice $k$, that population is strongly excited; when instead it is $x^{I}$ that lies near that region, choice $k$ is strongly inhibited.

\begin{figure*}[b]
    \centering
    \captionsetup{justification=centering}
    \begin{minipage}[c]{0.9\linewidth}
        \centering
        \includegraphics[keepaspectratio,width=\linewidth]{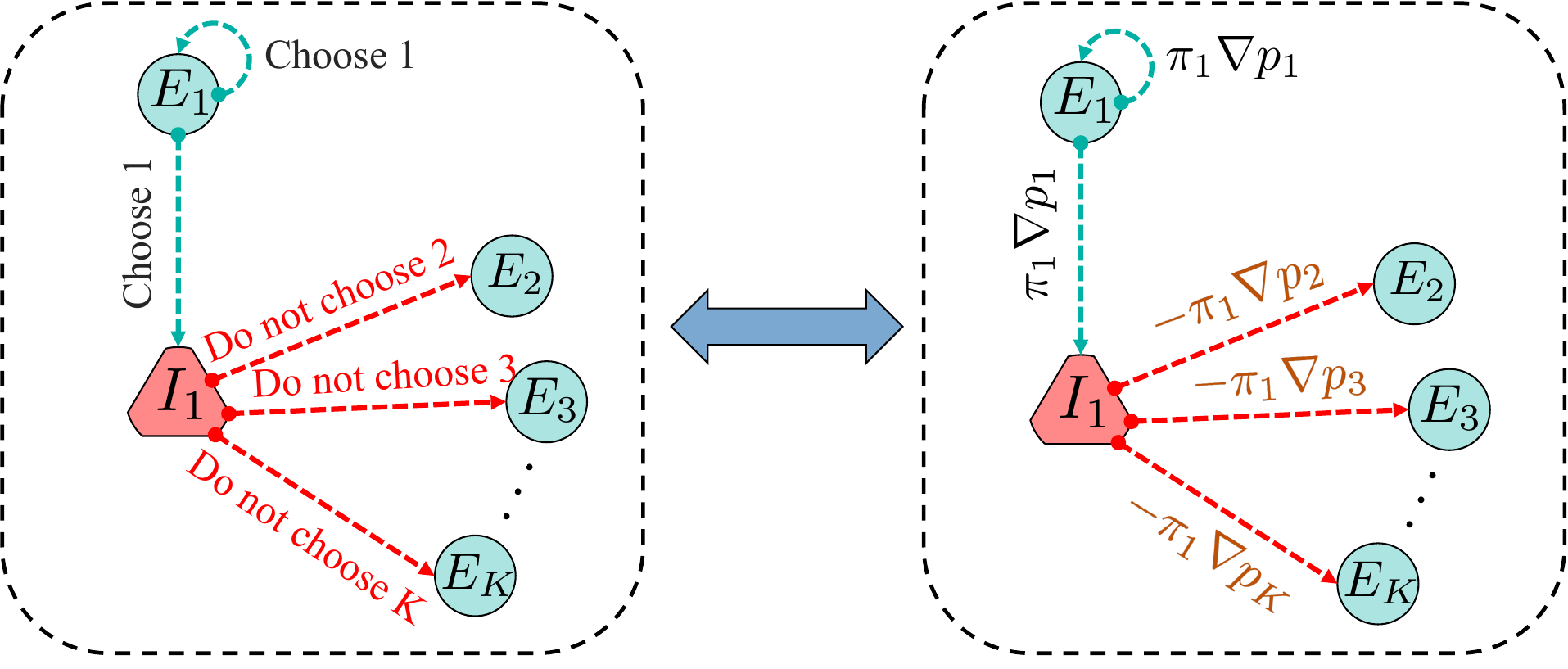}
        \subcaption{Opponent inhibition view 1}
        \label{fig:OI_model_view1}
    \end{minipage}
    \begin{minipage}[c]{0.9\linewidth}
        \centering
        \includegraphics[keepaspectratio,width=\linewidth]{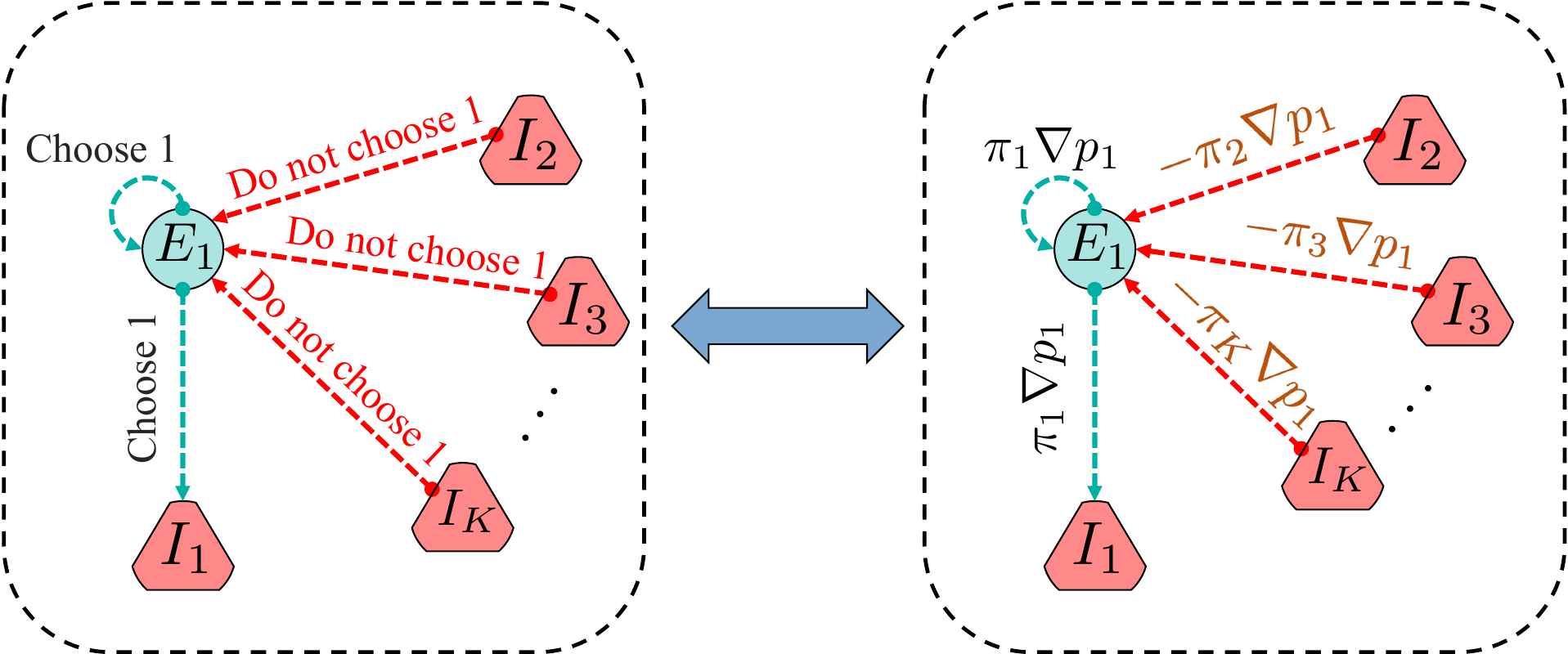}
        \subcaption{Opponent inhibition view 2}
        \label{fig:OI_model_view2}
    \end{minipage}
    \caption{Opponent inhibition model used as basis for the definition of the (RKL-based) excitatory and inhibitory drifts. Both views result in the same equations - or effect - and are therefore equivalent, resulting in the ``complementary" mixture $\bar{p}$.}
    \label{fig:opponent_inhib}
\end{figure*}

The effect of the inhibition can be formalized using a ``complementary'' mixture
\begin{align}
\bar{\pi}_k = \frac{1-\pi_k}{K-1}, \qquad
\bar{p}(x) = \sum_{k=1}^{K} \bar{\pi}_k\, p_k(x),
\end{align}
whose weights emphasize the non-chosen alternatives and are obtained through the combination of the inhibitory effect resulting from the alternative choices, i.e. $\sum\limits_{k=1}^{K} \pi_k \sum_{j\neq k} \nabla_{x} p_{j} = \sum\limits_{k=1}^{K} \sum_{j\neq k} \pi_j \nabla_{x} p_{k} = \sum\limits_{k=1}^{K} (1 - \pi_{k}) \nabla_{x} p_{k}$ as shown in figure~\ref{fig:opponent_inhib} (with the first term obtained from view 1 and the second one from view 2. Both lead to the same - third term - inhibitory contribution and result in the complementary mixture $\bar{p}$). Incorporating both excitation and inhibition yields the RKL-based OI drift
\begin{align}
dx_t
 &= \Bigl[
     w_E\,\nabla_{x}\ln p(x_t)
     - w_I\,\nabla_{x}\ln \bar{p}(-x_t)
   \Bigr] dt
   + \sqrt{2}\,dW_t,
\label{eq:OI_RKL_SDE_rewrite}
\end{align}
where $w_E$ and $w_I$ modulate the relative strengths of excitation and inhibition.

To generalize to the $\lambda$-KL objective, we can simply combine the drifts from two Wasserstein flows: one toward $p$ and one toward $\bar{p}$. Specifically, let $q^E_t$ and $q^I_t$ denote the variational densities corresponding to $p$ and $\bar{p}$. The excitatory and inhibitory drifts are then
\begin{align}
\mu_E(x_t)
 &= \bigl(1-\lambda + \lambda \tfrac{p(x_t)}{q^E_t(x_t)}\bigr)\,
     \nabla_{x} \ln p(x_t)
   + \lambda\bigl(1-\tfrac{p(x_t)}{q^E_t(x_t)}\bigr)\,
     \nabla_{x} \ln q^E_t(x_t), \\
\mu_I(x_t)
 &= \bigl(1-\lambda + \lambda \tfrac{\bar{p}(-x_t)}{q^I_t(-x_t)}\bigr)\,
     \nabla_{x} \ln \bar{p}(-x_t)
   + \lambda\bigl(1-\tfrac{\bar{p}(-x_t)}{q^I_t(-x_t)}\bigr)\,
     \nabla_{x} \ln q^I_t(-x_t),
\end{align}
and the full $\lambda$-KL OI dynamics are
\begin{align}
(w_E - w_I)^2 &= 1, \\
dx_t
 &= \bigl[ w_E\,\mu_E(x_t) - w_I\,\mu_I(x_t) \bigr] dt
    + \sqrt{2}\,dW_t.
\end{align}
In the simulations reported in the main paper, we set $w_E = 2$, and take $q^E_t$ and $q^I_t$ to be the static FKL minimizers for their respective mixtures $p$ and $\bar{p}$. The relation $(w_E - w_I)^2 = 1$ ensures that for the RKL Langevin dynamics, when both mixtures are equal $p = \bar{p}$, i.e. equally valued options, the SDE's drift reduces to $\nabla_{x} \ln p$.

\newpage

\section{Supplementary figures}

\begin{figure*}[b]
    \centering
    \includegraphics[keepaspectratio=true,width=\linewidth]{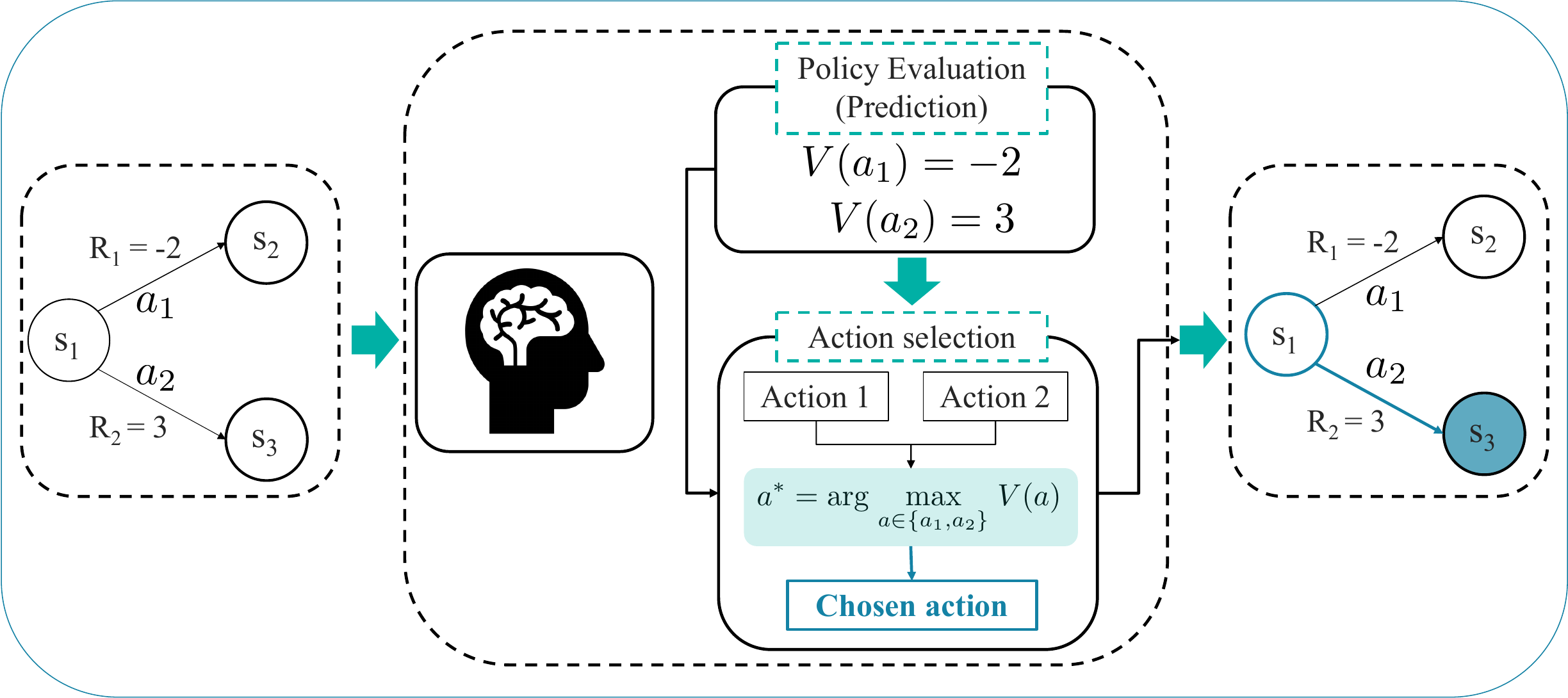}
    \caption{The two steps of conventional RL: policy evaluation (value learning or estimation) and policy improvement (action selection) with max operator.}
    \label{fig:rl}
\end{figure*}

\begin{figure*}[b]
    \centering
    \includegraphics[width=\linewidth]{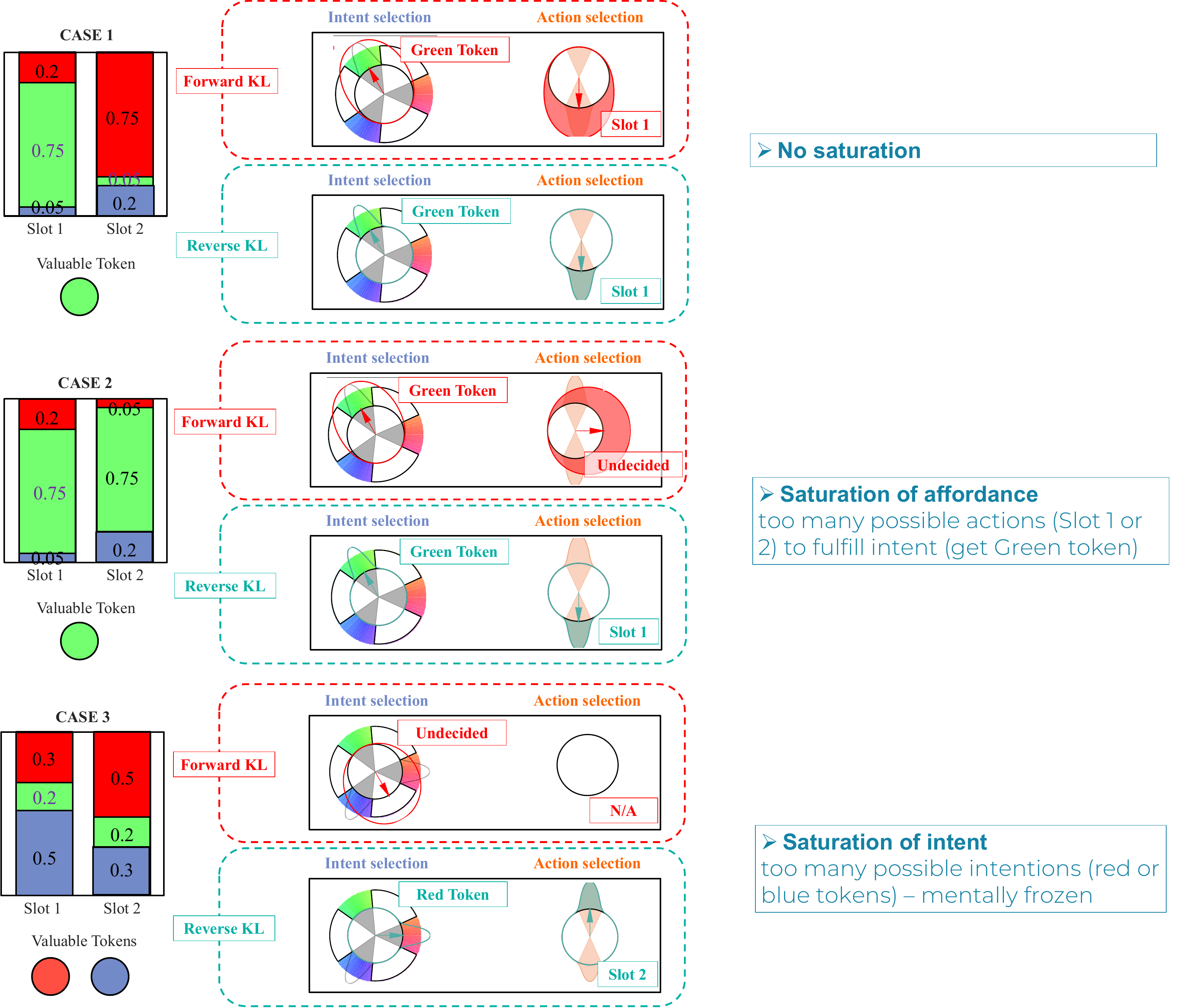}
    \caption{Static model simulations on the grid world and RGB token task. Actions and intents are represented as directions on a circular space. Shaded regions (\textcolor{red}{FKL} and \textcolor{mylightgreen}{RKL}) show selection frequencies; arrows mark mean selections. FKL and RKL behave similarly when a single option dominates, but FKL becomes indecisive under multimodal option structures, reflecting its mode-covering tendency.}
    \label{fig:rgb_static_res}
\end{figure*}

\begin{figure*}[b]
    \centering
    \includegraphics[width=\linewidth]{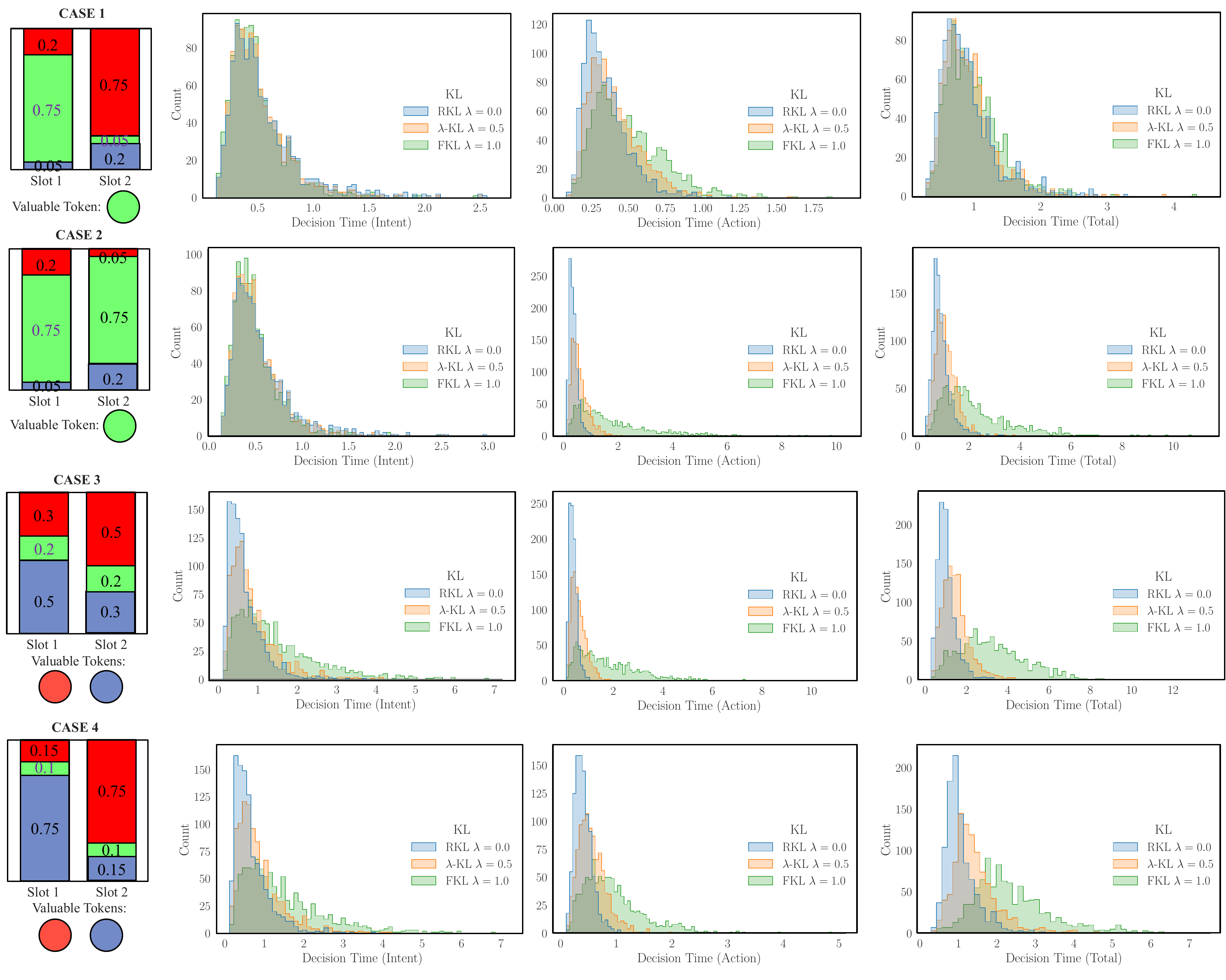}
    \caption{Decision time distributions across cases for three values of $\lambda$. Increasing $\lambda$ (approaching FKL) produces heavier tails, indicating more frequent long-latency decisions under ambiguous or multi-option conditions.}
    \label{fig:rgb_res}
\end{figure*}

\begin{figure*}[b]
    \centering
    \includegraphics[width=\linewidth]{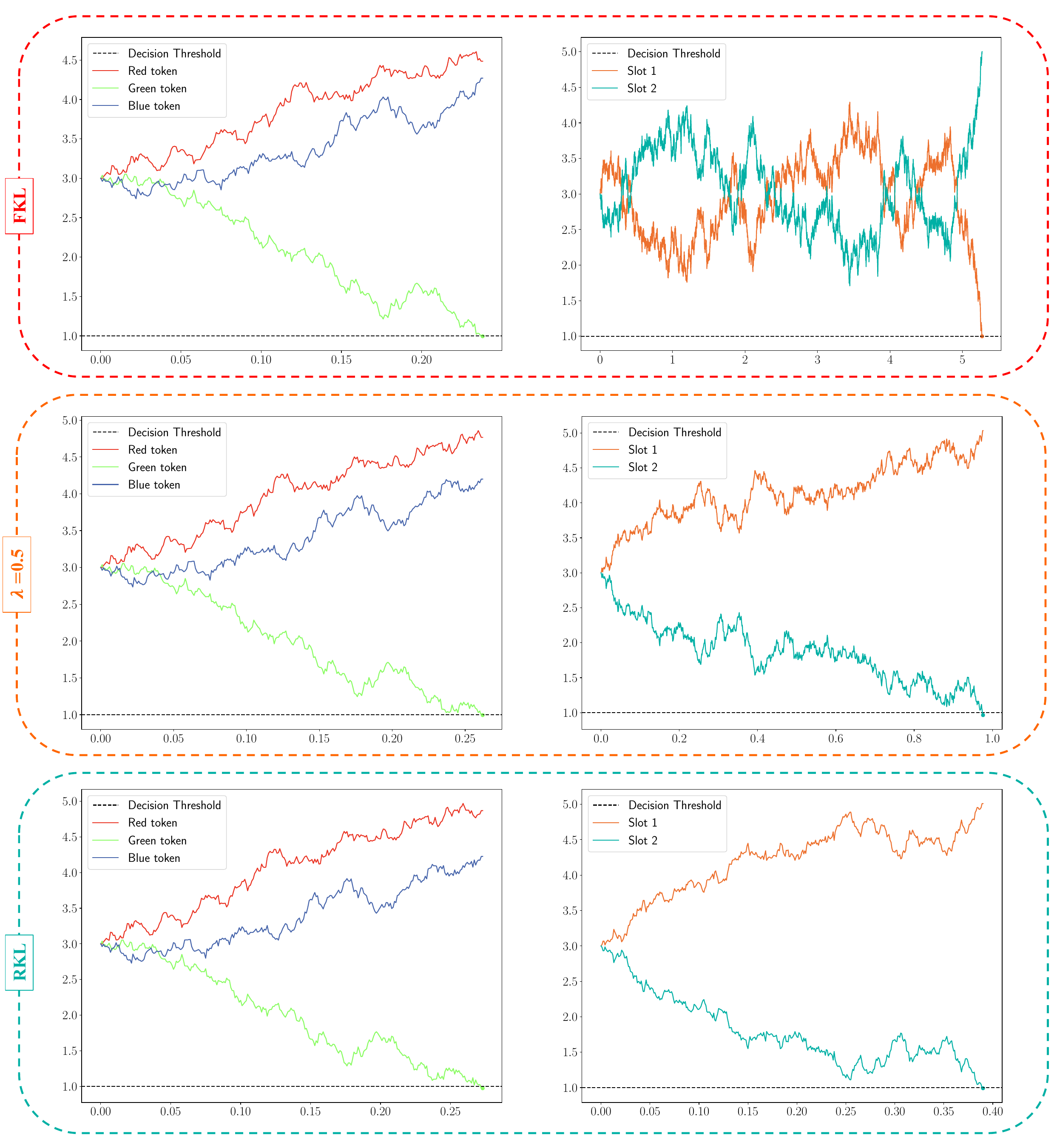}
    \caption{Example drift-diffusion trajectories for Case~2. Intent selection takes comparable time among models due to a single optimal intent. Action selection differs: RKL reaches a threshold rapidly, whereas FKL remains near the neutral zone for extended periods before committing.}
    \label{fig:rgb_ddm_res}
\end{figure*}

\end{appendices}

\end{document}